\documentclass[12pt]{article}
\usepackage{epsf}

\begin{document}
\setlength{\parskip}{2ex} \setlength{\parindent}{0em}
\setlength{\baselineskip}{3ex}
\newcommand{\onefigure}[2]{\begin{figure}[htbp]
         \caption{\small #2\label{#1}(#1)}
         \end{figure}}
\newcommand{\onefigurenocap}[1]{\begin{figure}[h]
         \begin{center}\leavevmode\epsfbox{#1.eps}\end{center}
         \end{figure}}
\renewcommand{\onefigure}[2]{\begin{figure}[htbp]
         \begin{center}\leavevmode\epsfbox{#1.eps}\end{center}
         \caption{\small #2\label{#1}}
         \end{figure}}
\newcommand{\comment}[1]{}
\newcommand{\myref}[1]{(\ref{#1})}
\newcommand{\secref}[1]{sec.~\protect\ref{#1}}
\newcommand{\figref}[1]{Fig.~\protect\ref{#1}}
\newcommand{\mathbold}[1]{\mbox{\boldmath $\bf#1$}}
\newcommand{\mJ}{\mathbold{J}}
\newcommand{\momega}{\mathbold{\omega}}
\newcommand{\bz}{{\bf z}}
\def\bbbz{{\sf Z\!\!\!Z}}
\newcommand{\PP}{\mbox{I}\!\mbox{P}}
\newcommand{\bbbc}{\mbox{C}\!\!\!\mbox{I}}
\def\sl2z{SL(2,\bbbz)}
\newcommand{\bbbq}{I\!\!Q}
\newcommand{\be}{\begin{equation}}
\newcommand{\ee}{\end{equation}}
\newcommand{\bea}{\begin{eqnarray}}
\newcommand{\eea}{\end{eqnarray}}
\newcommand{\nn}{\nonumber}
\newcommand{\unit}{1\!\!1}
\newcommand{\half}{\frac{1}{2}}
\newcommand{\shalf}{\mbox{$\half$}}
\newcommand{\transform}[1]{
   \stackrel{#1}{-\hspace{-1.2ex}-\hspace{-1.2ex}\longrightarrow}}
\newcommand{\inter}[2]{\null^{\#}(#1\cdot#2)}
\newcommand{\lprod}[2]{\vec{#1}\cdot\vec{#2}}
\newcommand{\mult}[1]{{\cal N}(#1)}
\newcommand{\Bn}{{\cal B}_N}
\newcommand{\B}{{\cal B}}
\newcommand{\Beight}{{\cal B}_8}
\newcommand{\Bnine}{{\cal B}_9}
\newcommand{\Eman}{\widehat{\cal E}_N}
\newcommand{\C}{{\cal C}}
\newcommand{\Q}{Q\!\!\!Q}
\newcommand{\comp}{C\!\!\!C}

\newdimen\tableauside\tableauside=1.0ex
\newdimen\tableaurule\tableaurule=0.4pt
\newdimen\tableaustep
\def\phantomhrule#1{\hbox{\vbox to0pt{\hrule height\tableaurule width#1\vss}}}
\def\phantomvrule#1{\vbox{\hbox to0pt{\vrule width\tableaurule height#1\hss}}}
\def\sqr{\vbox{%
  \phantomhrule\tableaustep
  \hbox{\phantomvrule\tableaustep\kern\tableaustep\phantomvrule\tableaustep}%
  \hbox{\vbox{\phantomhrule\tableauside}\kern-\tableaurule}}}
\def\squares#1{\hbox{\count0=#1\noindent\loop\sqr
  \advance\count0 by-1 \ifnum\count0>0\repeat}}
\def\tableau#1{\vcenter{\offinterlineskip
  \tableaustep=\tableauside\advance\tableaustep by-\tableaurule
  \kern\normallineskip\hbox
    {\kern\normallineskip\vbox
      {\gettableau#1 0 }%
     \kern\normallineskip\kern\tableaurule}%
  \kern\normallineskip\kern\tableaurule}}
\def\gettableau#1 {\ifnum#1=0\let\next=\null\else
  \squares{#1}\let\next=\gettableau\fi\next}

\tableauside=1.0ex
\tableaurule=0.4pt

\def\IE{\relax{\rm I\kern-.18em E}}
\def\IP{\relax{\rm I\kern-.18em P}}

\noindent
\begin{titlepage}

\begin{center}
\today \hfill UTTG-08-02\\ \hfill hep-th/0207114\\ \vskip 1cm
{\large {\bf  All Genus Topological String Amplitudes and 5-brane Webs 
as Feynman Diagrams}} \vskip 2cm
{ Amer Iqbal}\\ \vskip 0.5cm

{Theory Group, Department of Physics\\
University of Texas at Austin,\\
Austin, TX 78712, U.S.A.\\}

\end{center}

\begin{abstract}
A conjecture for computing all genus topological closed string
amplitudes on toric local Calabi-Yau threefolds, by interpreting the
associated 5-brane web as a Feynman diagram, is given. A propagator and
three point vertex is defined which allows us to write down the
amplitude associated with 5-brane web. We verify the conjecture
that this amplitude is equal to the closed string partition function
by computing integer invariants for resolved conifold and certain
curves of low degree in local del Pezzo surfaces, local Hirzebruch
surfaces and their various blowups.
\end{abstract}
\end{titlepage}
\newpage

\thispagestyle{empty}

\pagenumbering{arabic}

\section{Introduction}
Topological string amplitudes have been of interest to physicists and
mathematicians for a long time. For physicists these amplitudes compute
terms in the low energy effective action of the theory obtained by
compactifying IIA sting theory on a Calabi-Yau threefold \cite{kodaira-spencer}. They
are of interest to  mathematicians since they are the generating
functions of Gromov-Witten invariants \cite{Gromov,GW} of holomorphic curves in
Calabi-Yau threefold \cite{AM}.

For the case of toric Calabi-Yau threefolds various techniques are
available for computing these amplitudes. Localization can be used
for a direct calculations in the A-model \cite{ABmodel}.  This was discussed in detail
in \cite{CKYZ}
for genus zero case using mirror symmetry. Higher genus calculations, using localization, for 
the local $\PP^{2}$ case were carried out in \cite{KZ}.
B-model calculation, although tedious, provide
another way of calculating these amplitudes up to certain unknown
constants which can sometimes be fixed from their behavior at the
singular points \cite{GhV} in the moduli space and using the hidden integrality
properties of the invariants \cite{KKV}.

Integer invariants of holomorphic curves in Calabi-Yau
threefolds were defined in \cite{GV2} and it was shown that topological
closed string amplitudes when interpreted from the target space point
of view are the generating functions of these integer invariants,
\bea
F(\omega)=\sum_{g=0}^{\infty}g_{s}^{2g-2}F_{g}(\omega)=
\sum_{m=1}^{\infty}\sum_{\beta\in H_{2}(X,\bbbz)}\sum_{r=0}^{\infty}
\frac{N^{r}_{\beta}}{m}(2\mbox{sin}(m\frac{g_{s}}{2}))^{2r-2}\,e^{-m\beta\cdot\omega}\,.
\label{closed}
\eea
Where $X$ is the Calabi-Yau with K\"ahler form $\omega$ and
$N^{r}_{\beta}\in \bbbz$ are the invariants.
Topological open string amplitudes were reformulated in terms of open string
integer invariants in \cite{OV1} by interpreting them from the target space point of
view.

A completely new way of obtaining the topological closed string
amplitudes from the Chern-Simons theory was developed in \cite{GV1}
using geometric transition: deformation versus resolution of the
singularity. The case of blown up conifold was discussed in detail and
all genus closed string amplitude of the blown up conifold was derived
from the partition function of the Chern-Simons theory on $S^{3}$.
Recently this open-closed duality, Chern-Simons theory being the
theory of topological open strings, was proved from the world-sheet
point of view \cite{worldsheet}. The fact that both open and closed
string invariants are related to knot invariants in a geometry
obtained via geometric transition has been checked for many cases
\cite{LM,RS,LMV,SV,MV,DFG}.

Closed string invariants for more complicated geometries, such as
local toric del Pezzo surfaces, have been calculated from the
Chern-Simons theory using geometric transition \cite{DFG,AMV}. Unlike
the case of the resolved conifold, in these cases the Chern-Simons
theory on the space obtained after geometric transition gets
corrections from the holomorphic curves with boundaries on the
3-cycles \cite{WCS}.  This open-closed duality, Chern-Simons theory
being the theory of open strings \cite{WCS}, has been checked in many
cases and is by far the easiest method for calculating the
Gromov-Witten invariants and check their hidden integrality properties
as formulated in terms of integer invariants \cite{GV2}

It is well known that knot invariants in the Chern-Simons theory can
be calculated using the WZW theory. In this note we conjecture that
using the 5-brane web description of toric local Calabi-Yau threefolds
\cite{LV} the topological closed string partition function can be
written directly using the propagator and vertex defined by states and
operators in the large $N$ WZW theory. Since the 5-brane web naturally
gives rise to a Riemann surface the topological closed string
partition function is an amplitude associated with this Riemann
surface.  The conjecture is motivated by the lattice model
interpretation of the topological closed string amplitudes given in
\cite{AMV}. It follows from their discussion once we recognize that
the ``three point vertex'' is more basic than the four point vertex of
the lattice model \cite{AMV} and that all 5-brane web diagrams can be
constructed from (1,0) 5-brane (``the propagator'') , three point
vertex given by (1,0), (0,1) (1,1) 5-branes and their $\sl2z$
transforms. We verify the idea by computing integer invariants, from
this 5-brane web amplitude, for various geometries.

Using the 5-brane/7-brane description of certain local Calabi-Yau
threefolds, developed in \cite{DHIK} we conjecture the form of closed
string partition function for some non-toric threefolds. From the
5-brane/7-brane description of toric local Calabi-Yau threefolds it
appears that closed string partition function is given by particular
Feynman diagrams in the vacuum to vacuum amplitude. The case of non-toric
threefolds will be discussed in more detail elsewhere\cite{PI}.

This paper is organized as follows. In section two we discuss briefly
the 5-brane web description of toric local Calabi-Yau threefolds. We
also discuss, in this section, the Riemann surface coming from the
5-brane web when IIB string theory is lifted to M-theory after
compactification on a circle. In section two we define the propagator
and the three point vertex using the states and the operators of the
WZW theory. We also discuss the matrix elements and and their
computation which will be needed later to work out the integer
invariants. The matrix elements we will need are related to the Hopf
link with linking number plus one and therefore the discussion of
matrix elements is mostly based on section three of \cite{AMV}.  In
section three we compute integer invariants for a few curves of low
degree in various toric local Calabi-Yau threefolds with a single
compact divisor from the 5-brane web description. A more complete
account of these invariants for curves of higher degree in these
geometries will be given elsewhere \cite{PI}.

\section{Local threefolds and 5-brane webs and the Riemann surface}
In this section we discuss toric local Calabi-Yau threefolds, their
dual description involving $(p,q)$ 5-branes of type IIB string theory and
the Riemann surface associated with the web.

\subsection{CY and 5-brane web}
The relation between local toric Calabi-Yau spaces and 5-brane webs of
type IIB string theory follows from the basic duality between M-theory
on $T^{2}$ and type IIB string theory on $S^{1}$. Consider as an
example the resolved conifold which has a toric description given by
the following equation \cite{witten}, \bea
&&|X_{1}|^{2}+|X_{2}|^{2}-|X_{3}|^{2}-|X_{4}|^{2}=r,\\ \nn
&&(X_{1},X_{2},X_{3},X_{4}) =
(e^{i\theta}\,X_{1},e^{i\theta}\,X_{2},e^{-i\theta}\,X_{3},e^{-i\theta}\,X_{4})\,.
\eea \onefigure{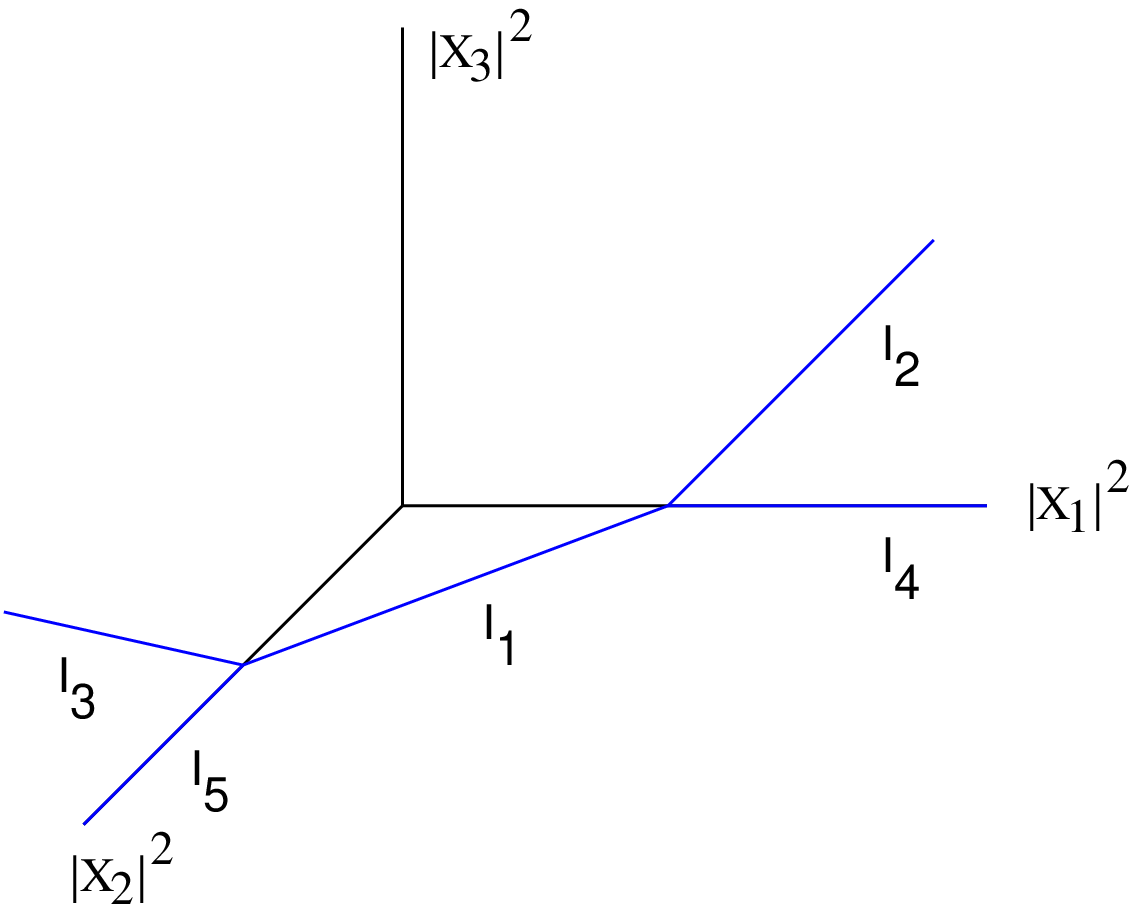}{The resolved conifold. Line segments
$I_{a}$ represent different collapsing $U(1)$'s.}  At a point, where
none of $X_{i}$ are zero we have a $T^{4}/U(1)$.  Where the $U(1)$
acts with charges $(1,1,-1-1)$ on the phases of $X_{a}$ which form the
$T^{4}$. On different line segments $I_{a}$, shown in
\figref{toric-conifold}, two different $X_{a}$ become zero and
therefore we are left with a single $U(1)$ after modding out. If we
consider the $\mbox{R}^{2}$ orthogonal to the line passing through the
origin and making equal angle with all three axis then we get a
$T^{2}$ fibration over this plane with $T^{2}$ degenerating as shown
in \figref{toric-conifold2}{}.  \onefigure{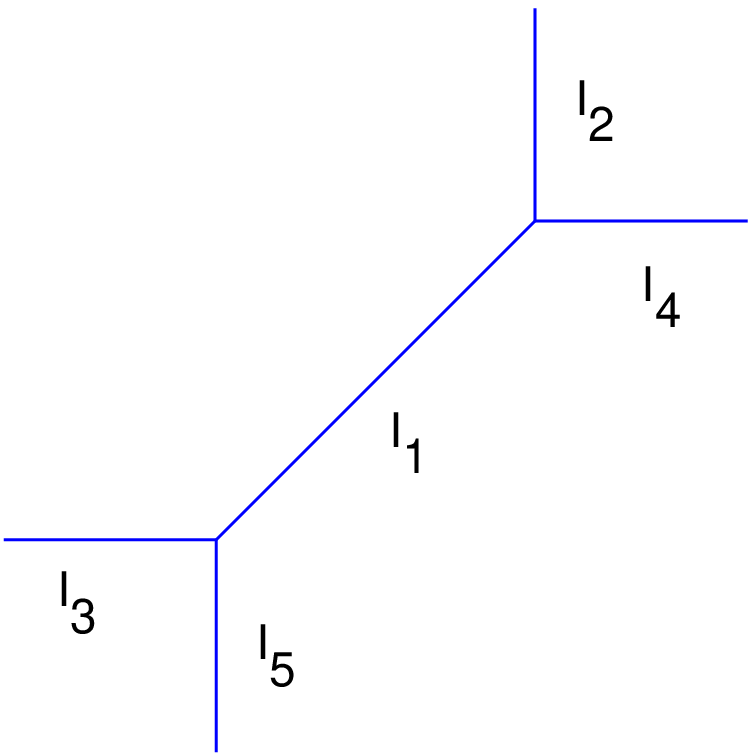}{Projection
of the resolved conifold.} On this plane we can choose the directions
such that $(p,q)$ cycle degenerates along a line segment oriented in
the $(p,q)$ direction. Now if we apply M-theory/IIB duality
adiabatically we see that the $T^{2}$ disappears and its information
is encoded in the branes present along the locus of degeneration. The
line segment directed in the $(p,q)$ direction , along which $(p,q)$
cycle was degenerating, now has $(p,q)$ 5-brane of IIB string theory.

The 5D ${\cal N}=1$ theory on the transverse space coming from
M-theory on the CY is dual to the 5D ${\cal N}=1$ theory on the
5-brane web. By compactifying one direction transverse to the conifold
we can go to type IIA string theory on the blown up
conifold. Compactifying one of the common directions of the 5-branes
we can go back to M-theory with 5-brane web becoming a single M5-brane
wrapped on a Riemann surface embedded in $\mbox{R}^{2}\times
T^{2}$. We will see that this Riemann surface plays the central role
in writing down the string partition function. This perhaps is related
to the results of \cite{bonelli, dijkgraaf1} where relation between
string partition functions and M5-brane theory was explored. The Riemann
surface projects to $\mbox{R}^{2}$ as a thickened 5-brane web.  The
${\cal N}=2$ theory on the transverse space obtained by compactifying
type IIA string theory on a local Calabi-Yau threefolds has a dual
description as the theory on an M5-brane wrapped on the Riemann
surface associated with the corresponding web.
\onefigure{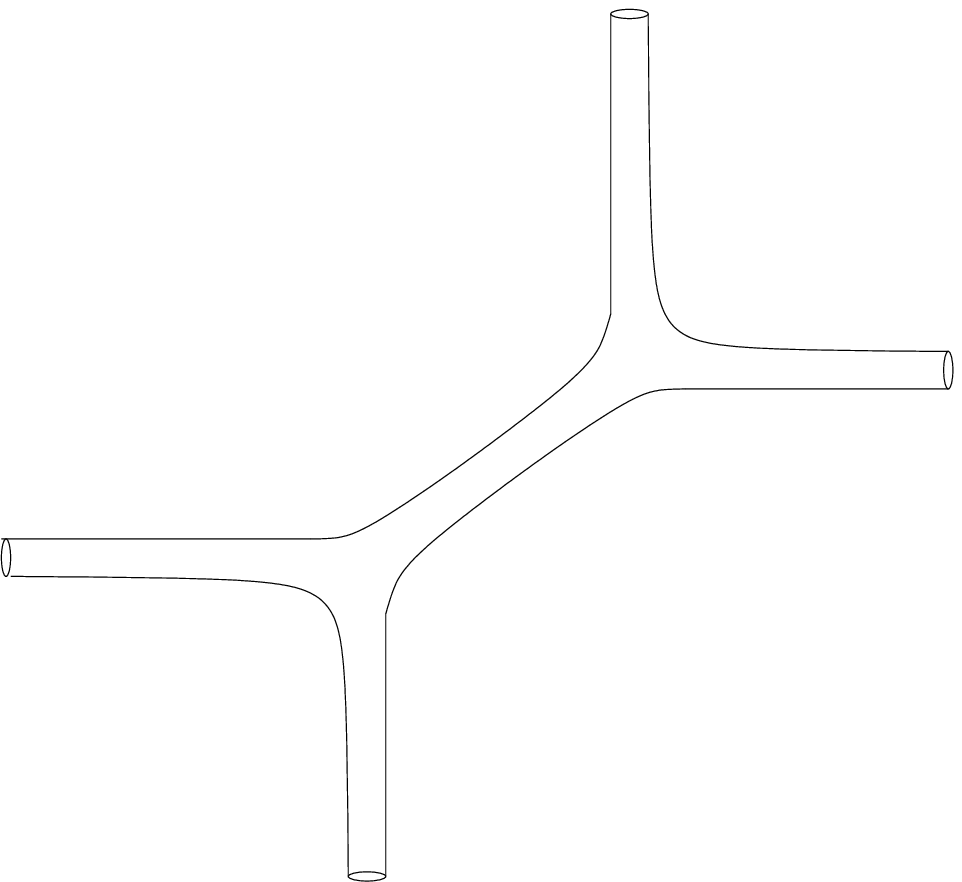}{The Riemann surface associated with the
5-brane web dual to the resolved conifold.}  For toric local
Calabi-Yau this Riemann surface can easily be obtained from the toric
data and plays an important role in constructing the mirror Calabi-Yau
threefold. If the Riemann surface associated with web is given by
$f(e^{u},e^{v})=0$ (u,v being coordinates on $R^{2}\times T^{2}$) then
the mirror Calabi-Yau threefold is given by \bea
f(e^{u},e^{v})=xy\,,\,\,x,y\in \bbbc.  \eea

\section{5-brane webs as Feynman diagrams}

5-brane webs provide an interesting way of encoding the geometry of
the dual Calabi-Yau threefold. The curves in the CY which give rise to
BPS states of the ${\cal N}=1$ 5D theory correspond to $(p,q)$ strings ending
on the $(p,q)$ 5-branes of the web. From this web description of
curves it is easy to compute the intersection number of curves and
also the genus of a curve using the fact that the web of $(p,q)$
strings should be trivalent \cite{AHK, LV, KR2}.

The most basic web is the one composed of the $(1,0)$, $(0,1)$ and the
$(1,1)$ 5-brane. This web corresponds to the threefold $\bbbc^{3}$ as
can be seen by taking the limit in which the K\"ahler parameter of the
$\PP^{1}$ of the resolved conifold goes to infinity. Since we are in
type IIB string theory which has $\sl2z$ symmetry, therefore their is
no unique web corresponding to a local toric Calabi-Yau threefold. By an $\sl2z$
transformation $\pmatrix{p &r \cr q & s}$ we can map the $(1,0)$,
$(0,1)$ and $(1,1)$ web to $(p,q)$, $(r,s)$ and $(p+q,r+s)$ web. The
invariant aspect of this transformed web is the intersection
number of the external charges, \bea \mbox{det}\pmatrix{p& r\cr q&s}=1\,.  \eea
If we have a web with three 5-brane meeting a point such that the
intersection number of any two of them is not $\pm 1$ then it is
possible to resolve into a web with more than one trivalent vertices
such that the intersection numbers at both vertices are $\pm 1$. The
fact that the intersection number is not $\pm 1$ is reflection of the
fact that the corresponding CY has a singularity \cite{LV}. The
resolution of the singularity of the Calabi-Yau is then reflected in
the dual web diagram as the appearance of more vertices. A simple
example of this is given by the web diagram of the Calabi-Yau which is
the total space of ${\cal O}(-3)$ bundle over $\PP^{2}$. If the
$\PP^{2}$ is shrunk then the CY has a singularity and the web diagram
of the singular CY consists of $(-1,1)$, $(2,1)$ and $(-1,-2)$
branes. The intersection numbers in this case are $\pm 3$ (depending
on the orientation) and therefore the web can be resolved as shown in
\figref{resolution}.  \onefigure{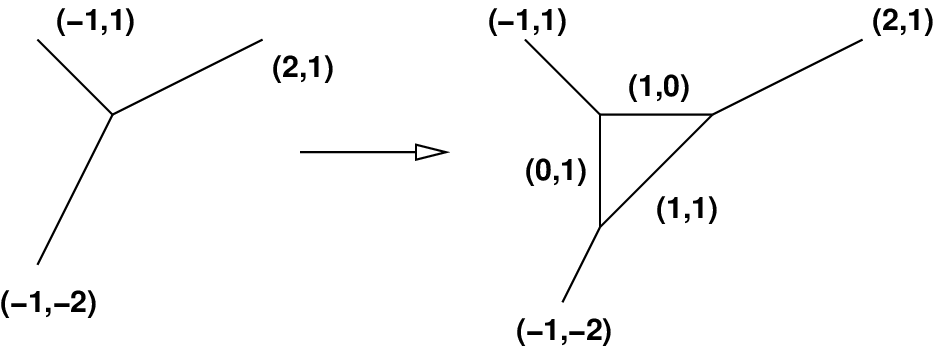}{Resolution of $\bbbc^{3}/\bbbz_{3}$.}
 It is easy to check
that the resolved web has intersection numbers $\pm 1$ at each vertex
and therefore each vertex is an $\sl2z$ transform of the basic
$(1,0)$, $(0,1)$ and $(1,1)$ vertex. Thus we can decompose any web
into pieces which are $\sl2z$ transform of the basic three point
vertex and the $(1,0)$ 5-brane as shown in \figref{pieces} for the
case of resolved conifold and local $\PP^{2}$.  \onefigure{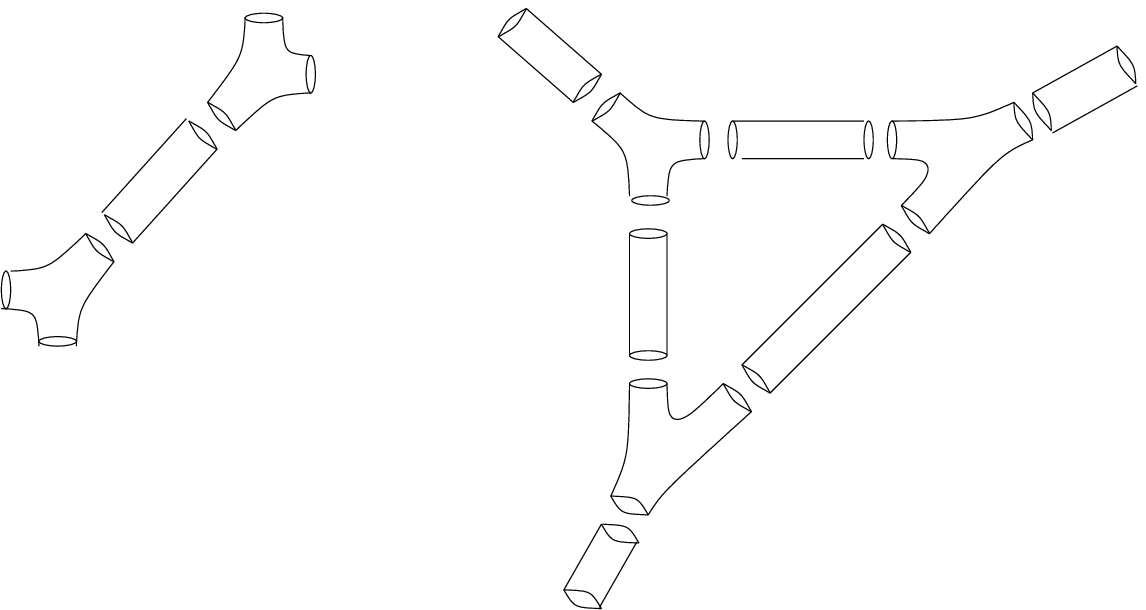}{a) Resolved conifold, b)local 
$\PP^{2}$. The Riemann surface associated with the web, and the web itself, can be constructed
from two basic objects.}

In \cite{AMV} a lattice model interpretation of topological closed
string amplitudes was discussed and it was suggested that with a web
configuration in which all three point vertices can be converted into
four point vertices (in the CY context this corresponds to shrinking
$\PP^{1}$'s with normal bundle ${\cal O}(-1)\oplus {\cal O}(-1)$)
matrix element $\langle \bar{R_{3}},\bar{R_{4}}|V|R_{1},R_{2}\rangle$ should be
associated with the four point vertices and propagators $e^{-l_{R}r}$
with the edges. Where $|R>$ are the states of the $SU(N)$ WZW theory
which are labeled by the highest weight representations $R$ at level
$k+N$ and $l_{R}$ is the first Casimir of the representation $R$ equal
to the number of boxes in the Young-Tableaux of $R$. $r$ is the length
of the edge and is the K\"ahler parameter of the associated rational
curve in the dual Calabi-Yau. In the case of an external edge
we have $r\mapsto \infty$ and therefore the state associated with the
external legs is the vacuum state $|0\rangle$. As discussed in \cite{AMV} the state
$|R_{1},R_{2}\rangle$ is given by the fusion coefficients,
\bea
|R_{1},R_{2}\rangle =\sum_{R}N^{R}_{R_{1}R_{2}}\,|R\rangle\,,
\eea
where the fusion coefficient are given in terms of the modular $S$ matrix,
\bea
N^{R}_{R_{1}R_{2}}=\sum_{R'}\frac{S_{R'R_{1}}S_{R'R_{2}}S^{-1}_{R'R}}{S_{0R'}}\,.
\eea

We have seen that three point vertex is more basic than the four point
vertex and therefore it is should be possible to associate with it a
matrix element $\langle \bar{R}_{3}|{\cal O}| R_{2},0\rangle$, with
${\cal O}$ being some appropriate operator and $|0\rangle $
,$|R_{2}\rangle $ states on the incoming circles and $|R_{3}\rangle$
the state on the outgoing circle as shown in \figref{vertex}. We
always consider the vertices in which one of the boundary circles goes
of to infinity so that the state on it is the vacuum state.
\onefigure{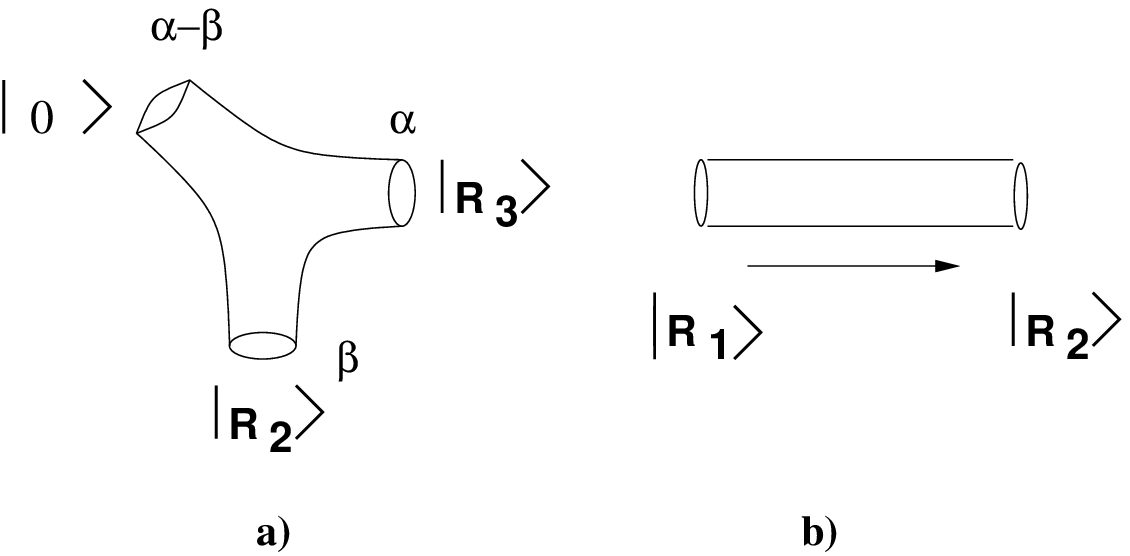}{a) The vertex, b) the propagator.} We take the
amplitude for this vertex to be \footnote{I am grateful to Cumrun Vafa
for pointing out a mistake in the definition of the vertex in an
earlier version of the preprint.}\bea
V(\alpha,\beta;\alpha-\beta):=\langle
\bar{R_{3}}|S^{-1}|R_{2},0\rangle\,, \eea where the operator ${\cal
O}=S^{-1}$ has been chosen such that the corresponding $\sl2z$ matrix
takes the incoming circle $\beta$ into an outgoing circle
$\alpha$.\footnote{The Riemann surface is embedded in $\mbox{R}^{2}\times T^{2}$
 and $\alpha,\beta$ are the basis of $H_{1}(T^{2},\bbbz)$ such
that $\alpha\cdot\beta=1$.}  The importance of the three point vertex has also been noted
in \cite{paper1} where a vertex with all three circles having
non-trivial state has been worked out.

An asymmetrical choice between $\beta$ and
$\alpha-\beta$ was made here. Of the two incoming cycles we choose the
one which is followed by the outgoing cycle in the counter clock-wise
rotation on the vertex.  We will carefully choose our web diagrams
from among all $\sl2z$ transforms so that the operators at the
vertices are of the form $T^{m}S^{-1}T^{k}$ for $m,k\in \bbbz$. We
will see that this is possible for all local del Pezzo surfaces, local
Hirzebruch surfaces and their various blowups.  This choice is made so
that we can use the method given in \cite{AMV} to calculate the matrix
elements $\langle \bar{R}|T^{m}S^{-1}T^{k}|R'\rangle$.  For the
propagator which is cylinder of length $r$ with states $|R_{1}\rangle
$ and $|R_{2}\rangle $ at the two boundary circles
$J_{R_{1}},J_{R_{2}}\in H_{1}(T^{2},\bbbz)$ we choose, \bea
P(J_{R_{1}},J_{R_{2}}):=e^{-l_{R_{1}}r}\delta_{R_{1}R_{2}}\,.  \eea

Although not all web diagrams would allow operators at the vertices of
the form $T^{n}S^{-1}T^{m}$ simultaneously we will restrict ourselves
to geometries which do allow such vertices so that we can use the
results of \cite{AMV}. 

The states and the operators we used to define the propagator and the
vertices are that of the $U(N)$ WZW theory. It is well known that the
primary field of this theory are associated with highest weight
representations at level $k+N$
\cite{verlinde,jonespolynomial}.  A state $|R\rangle$ with weight
vector $\lambda_{R}$ has conformal dimension \bea
h_{R}=\frac{(\lambda_{R},\lambda_{R}+2\rho)}{2(k+N)} \eea where $\rho$ is one half the sum of positive
roots. These states
define an orthonormal basis of the Hilbert space of the Chern-Simons
theory on a manifold with boundary of genus one, $\langle
\bar{R}_{1}|R_{2}\rangle =\delta_{R_{1},R_{2}}$. As discussed in
\cite{verlinde} $\sl2z$ action on the boundary $T^{2}$ is implemented
on the Hilbert space using the operators $S$ and $T$ which form a
representation of the $\sl2z$, \bea
S^{2}=(ST)^{3}=C\,,\,\,\,\mbox{where}\,\,\,\,C_{R_{1},R_{2}}=\delta_{\bar{R_{1}},R_{2}}\,.
\eea In the basis chosen above the $T$ operator is diagonal with
eigenvalue $e^{2\pi i(h_{R}-c/24)}$ when acting on the state
$|R\rangle$. The $S$ operator is not diagonal and is given by
\bea
S_{R_{1},R_{2}}:=K
\sum_{w\in W}\epsilon(w) e^{-\frac{2\pi
i}{k+N}(w(\lambda_{R_{1}+\rho}),\lambda_{R_{2}}+\rho)}\,. \eea
Where $K$ only depends on $k$ and $N$ and $W$ is the Weyl group.
From the relation between the Hilbert space of the Chern-Simons theory
and the space of conformal blocks of the WZW theory
\cite{jonespolynomial} it follows that the matrix element
$W_{R_{1},R_{2}}:=\langle \bar{R}_{1}|S^{-1}|R_{2}\rangle$ is the link
invariant associated with the Hopf link of linking number $+1$ \cite{AMV}. This
essentially follows from the path integral representation of the
matrix element in the Chern-Simons theory. Since the action of $T$ is
diagonal it follows that \cite{AMV} \bea \langle
\bar{R_{1}}|T^{m}S^{-1}T^{n}|R_{2}\langle =\langle
\bar{R_{1}}|S^{-1}|R_{2}\rangle
\,(-1)^{ml_{R_{1}}+nlR_{2}}\,q^{\frac{1}{2}(m\kappa_{R_{2}}+n\kappa_{R_{1}})}\,,
\eea where $l_{R}$ as defined before is the number of boxes in the
Young-Tableaux of representation $R$ and $q=e^{\frac{2\pi
i}{k+N}}$. The integers $\kappa_{R}$ are defined in terms of the
number of rows $d(\mu^{R})$ in the Young Tableaux $\mu^{R}$ and number of boxes in a
a given row $\mu^{R}_{i}$ as \bea
\kappa_{R}=l_{R}+\sum_{i=1}^{d(\mu^{R})}\mu^{R}_{i}(\mu^{R}_{i}-2i).\eea
Thus we only need to calculate the matrix elements of $S^{-1}$ which can be done easily
using the results of \cite{ML,lukac,AMV} as we will explain now.

\underline{\bf $W_{R_{1},R_{2}}$:} The matrix element is a function of
$q$ and $\lambda:=q^{N}$. It is defined using the following q-numbers,
\bea
[x]:=q^{\frac{x}{2}}-q^{-\frac{x}{2}}\,,\,\,[x]_{\lambda}=\lambda^{\frac{1}{2}}q^{\frac{x}{2}}-\lambda^{-\frac{1}{2}}q^{\frac{x}{2}}\,,
\eea and a polynomial associated with each Young-Tableaux $\mu^{R}$
given by \bea
s_{\mu^{R}}(P(t))=\mbox{det}_{i,j}e_{\widehat{\mu}^{R}_{i}+j-i}\,,\,\,\,e_{0}=1,\,\,e_{k<0}=0\,.
\eea Where $\widehat{\mu}^{R}$ is the Young-Tableaux obtained by
interchanging the rows and the columns and $e_{i}$ are such that
$P(t)=1+\sum_{n=1}^{\infty}e_{n}t^{n}$. For example, \bea
s_{\tableau{1}}&=&e_{1}\,,\,s_{\tableau{2}}=e_{1}^{2}-e_{2}\,,\,\,s_{\tableau{1
1}}=e_{2}\,,\,\,\\ \nn
s_{\tableau{3}}&=&e_{3}-2e_{1}e_{2}+e_{1}^{3}\,,\,\,s_{\tableau{1 1
1}}=e_{3}\,,\,\,s_{\tableau{2 1}}=e_{3}-e_{1}e_{2}\,.  \eea The matrix
element $W_{R_{1},R_{2}}$ is given by \cite{ML,lukac,AMV} 
\bea
W_{R_{1},R_{2}}:=\mbox{dim}_{q}R_{1}\,(\lambda\,q)^{\frac{l_{R_{2}}}{2}}\,s_{\mu^{R_{2}}}(E_{\mu^{R_{1}}}(t))\,.
\eea 
Where 
\bea \mbox{dim}_{q}R:=\prod_{1\leq i<j\leq
d(\mu^{R})}\frac{[\mu_{i}-\mu_{j}+j-i]}{[j-i]}\prod_{i=1}^{d(\mu^{R})}\frac{\prod_{v=-i+1}^{\mu_{i}-i}[v]_{\lambda}}{\prod_{v=1}^{\mu_{i}}[v-i+d(\mu^{R})]}\,.
\eea 
and \bea
E_{\mu^{R}}(t)=(\prod_{j=1}^{d(\mu^{R})}\frac{1+q^{\mu_{j}-j}t}{1+q^{-j}t})\,(1+\sum_{n=1}^{\infty}
(\prod_{i=1}^{n}\frac{1-\lambda^{-1}q^{i-1}}{q^{i}-1})t^{n})\,.  \eea
We will see later that in the large $N$ limit keeping $q$ fixed we
will only need the leading order term, coefficient of
$\lambda^{\frac{l_{R_{1}}+l_{R_{2}}}{2}}$, in $W_{R_{1},R_{2}}$ which we denote by ${\cal W}_{R_{1},R_{2}}$. It is
easy to calculate the leading order term since the expression for
$E_{0}(t)$ simplifies in this limit and because the leading order term
in $\mbox{dim}_{q}R$ can be explicitly calculated and is given by
\cite{AMV} \bea q^{\kappa_{R}/4}\,\prod_{1\leq i<j\leq
d(\mu^{R})}\frac{[\mu_{i}-\mu_{j}+j-i]}{[j-i]}\prod_{i=1}^{d(R)}
\prod_{v=1}^{\mu_{i}}\frac{1}{[v-i+d(\mu^{R})]}\,.  \eea 

We list here ${\cal W}_{R_{1},R_{2}}$ for a few representations which we will need later 
for computing integer invariants.
\bea {\cal
W}_{\underbrace{\tableau{3}\cdots\tableau{1}}_{n}}&=&\frac{q^{n(n-1)/4}}{\prod_{k=1}^{n}(q^{k/2}-q^{-k/2})}\,,\\
\nn {\cal
W}_{A_{n}}&=&\frac{q^{-n(n-1)/4}}{\prod_{k=1}^{n}(q^{k/2}-q^{-k/2})}\,,\\
\nn {\cal W}_{\tableau{1},\tableau{1}}&
=&\frac{((q+q^{-1}-1)}{(q^{1/2}-q^{-1/2})^2})\,\\ \nn {\cal
W}_{\tableau{2},\tableau{1}}
&=&\frac{1}{(q^{1/2}-q^{-1/2})^3}\,\frac{(1-q^{2}+q^{3})}{1+q}\\
\nn {\cal W}_{\tableau{1
1},\tableau{1}}&=&\frac{1}{(q^{1/2}-q^{-1/2})^3}\,\frac{q^{-2}(1-q+q^3)}{1+q}\\
\nn {\cal
W}_{\tableau{3},\tableau{1}}&=&\frac{1}{(q^{1/2}-q^{-1/2})^4}\,\frac{q^{2}(1-q^3+q^4)}{(1+q)(1+q+q^2)}\\
\nn {\cal W}_{\tableau{2
1},\tableau{1}}&=&\frac{1}{(q^{1/2}-q^{-1/2})^4}\,\frac{1-q+q^2-q^3+q^4}{1+q+q^2}\,,\\
\nn {\cal W}_{\tableau{1 1
1},\tableau{1}}&=&\frac{1}{(q^{1/2}-q^{-1/2})^{4}}\,\frac{q^{1/2}(1-q+q^4)}{(1+q)(1+q+q^2)}\,\\
\nn {\cal
W}_{\tableau{2},\tableau{2}}&=&\frac{1}{(q^{1/2}-q^{-1/2})^4}\,\frac{q^{1/2}}{1+q}\,,\\
\nn {\cal W}_{\tableau{1
1},\tableau{2}}&=&\frac{1}{(q^{1/2}-q^{-1/2})^{4}}\,\frac{q^{-1}(1-q^{2}+q^4)}{(1+q)^2}\,,\\
\nn {\cal W}_{\tableau{1 1},\tableau{1
1}}&=&\frac{1}{(q^{1/2}-q^{-1/2})^4}\,\frac{q^{-4}(1-q+q^2+2q^3-q^5+q^6)}{(1+q)^2}\,.
\eea Where $A_{n}$ is the Young-Tableaux with a single column of $n$ rows and ${\cal W}_{R_{1}}={\cal W}_{R_{1},0}$.

\section{Resolved Conifold}
We begin by considering the case of the resolved conifold with the
following web diagram.  The operator at the right left vertex is
$S^{-1}$ which takes $\pmatrix{0\cr 1}$ to $\pmatrix{1\cr 0}$ and the
operator at the right vertex is also $S^{-1}$ since it takes
$\pmatrix{0\cr -1}$ to $\pmatrix{-1\cr 0}$. With vacuum states
associated with the boundary circles at infinity we get the amplitude
\onefigure{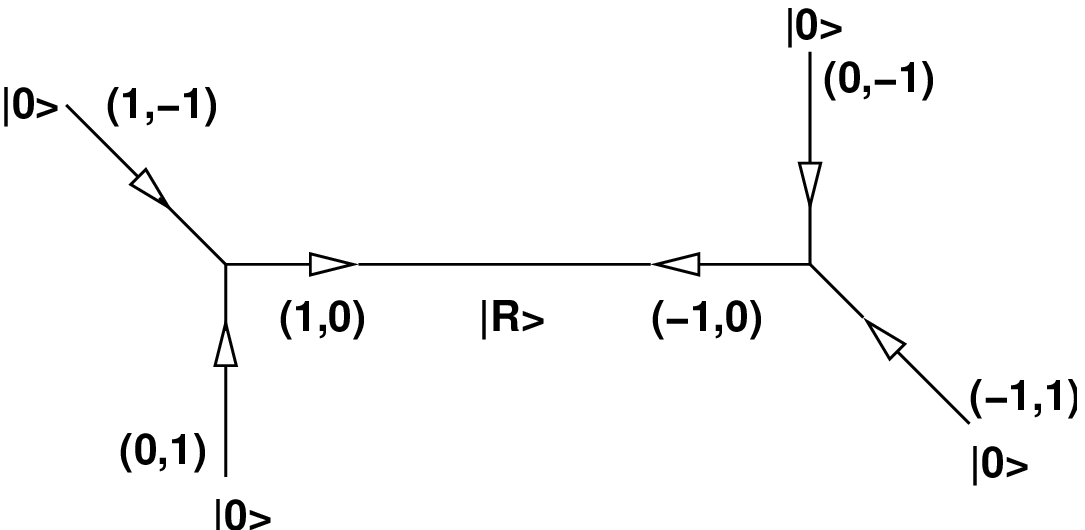}{5-brane web of the resolved conifold with vacuum states associated to the
boundary circles.} \bea Z_{conifold} &=&\sum_{R}\langle
0,0|S^{-1}|R\rangle e^{-l_{R}\,r}\langle
\bar{R}|S^{-1}|0,0\rangle\,,\\ \nn &=& \sum_{R}\langle
0|S^{-1}|R\rangle e^{-l_{R}\,r}\langle \bar{R}|S^{-1}|0\rangle\,.
\label{conifoldamplitude} \eea
We claim that the instanton corrections to the topological closed string amplitude on the resolved conifold $F^{(c)}_{inst}$ is given by
\bea
F^{(c)}_{inst}(T)=-\mbox{log}Z_{conifold}(r)\,,\,\,\,q=e^{ig_{s}}\,.
\eea
Where $T$ is the renormalized K\"ahler parameter. 
Thus $-\mbox{log}Z_{conifold}$ gives the instanton corrections to the closed
topological string amplitude when written in terms of the renormalized K\"ahler parameter
$T$. To determine the relation
between $r$ and $T$ we calculate
the term proportional to $e^{-r}$, \bea I_{1}&=&e^{-r}\langle
0|S^{-1}|R\rangle \langle \bar{R}|S^{-1}|0\rangle \,,\\ \nn
&=&e^{-r}\,W^{2}_{\tableau{1}}\,,\\ \nn
&=&e^{-r}\,(\frac{\lambda^{1/2}-\lambda^{-1/2}}{q^{1/2}-q^{-1/2}})^{2}\,.
\eea In the large $N$ limit $\lambda\mapsto \infty$ and therefore to get a finite non-vanishing result we
define $T$ as follows
\bea
r=T+\mbox{log}(\lambda)\,.
\eea
Then we can write Eq(\ref{conifoldamplitude}) as
\bea
e^{-F^{(c)}_{inst}}=\sum_{R}e^{-l_{R}T}\,{\cal W}^{2}_{R}(q)\,.
\label{conifoldamplitude2}
\eea
Where as defined in the last section before ${\cal W}_{R}(q)$ is the coefficient of $\lambda^{l_{R}/2}$ in
$W_{R}(q):=\langle 0|S^{-1}|R\rangle$.
To see this gives the correct closed string expansion we calculate a few terms. Denoting the term proportional to $e^{-nT}$ by $I_{n}$ we have,
\bea
\label{expansionxx}
I_{1}&:=& e^{-T}{\cal W}^{2}_{\tableau{1}}=e^{-T}\,\,\frac{1}{(q^{1/2}-q^{-1/2})^{2}}\\\nn
I_{2}&:=&e^{-2T}\{{\cal W}^{2}_{\tableau{2}}+{\cal W}^{2}_{\tableau{1 1}}\}=
e^{-2T}\,\,\frac{q+q^{-1}}{(q^{1/2}-q^{-1/2})^{4}(q^{1/2}+q^{-1/2})^2}\,,\\ \nn
I_{3}&:=&e^{-3T}\{{\cal W}^{2}_{\tableau{3}}+{\cal W}^{2}_{\tableau{2 1}}+{\cal W}^{2}_{\tableau{1 1 1}}\}=e^{-3T}
\frac{(q^{3}+q^{-3})+(q+q^{-1})+2}{(q^{1/2}-q^{-1/2})^{4}(q^{1/2}+q^{-1/2})^{2}(q^{3/2}-q^{-3/2})^2}\,,\\ \nn
I_{4}&:=&e^{-4T}\{{\cal W}^{2}_{\tableau{4}}+{\cal W}^{2}_{\tableau{3 1}}+{\cal W}^{2}_{\tableau{2 2}}+{\cal W}^{2}_{\tableau{2 1 1}}+{\cal W}^{2}_{\tableau{1 1 1 1}}\}\\ \nn
&=&e^{-4T}\frac{C(q)}{(q^{1/2}-q^{-1/2})^{6}(q^{1/2}+q^{-1/2})^{4}(q+q^{-1})^{2}(q^{3/2}-q^{-3/2})^{2}}\,.
\eea

Where
\bea
C(q):=(q^{6}+q^{-6})+(q^{4}+q^{-4})+2(q^{3}+q^{-3})+4(q^{2}+q^{-2})+2(q+q^{-1})+4\,.
\eea
Since the resolved conifold has a single $\PP^{1}$ 
from Eq(\ref{closed}) it follows that the closed string expansion for the conifold is
 given by \cite{GV2}
\bea
F^{(c)}_{inst}(T)=\sum_{n=1}^{\infty}\sum_{g=0}^{\infty}
\frac{N^{g}_{C}}{n}(2\mbox{sin}(ng_{s}/2))^{2g-2}\,e^{-nT}\,.
\eea
Comparing this with Eq(\ref{conifoldamplitude2}) using Eq(\ref{expansionxx}) we get
\bea
-\mbox{log}Z_{conifold}&=&\mbox{log}(1+I_{1}+I_{2}+I_{3}+I_{4}+\cdots)\,,\\ \nn
&=& I_{1}+\{I_{2}-\frac{1}{2}I_{1}^{2}\}+\{I_{3}-I_{2}I_{1}+\frac{1}{3}I_{1}^{3}\}+
\{I_{4}-I_{3}I_{1}+I_{2}I_{1}^{2}-\frac{1}{2}I_{2}^{2}-\frac{1}{4}I_{1}^{4}\}+\cdots\,,\\ \nn
&=& \frac{e^{-T}}{(2\mbox{sin}(g_{s}/2))^{2}}+\frac{e^{-2T}}{2(2\mbox{sin}(g_{s}))^{2}}+\frac{e^{-3T}}{3(2\mbox{sin}(3g_{s}/2))^{2}}+\frac{e^{-4T}}{4(2\mbox{sin}(2g_{s}))^{2}}+\cdots\,.
\eea
Which agrees with the expected closed string expansion.

\section{Calabi-Yau with single compact divisor}
In this section we will discuss non-compact toric Calabi-Yau threefolds which
have a single compact divisor. These Calabi-Yau manifolds are the
total space of the canonical line bundle over a toric Fano
or toric almost Fano surface. We will consider the case of blowups of $\PP^{2}$ and
the blowups of first and second Hirzebruch surfaces. We denote the blow up of $\PP^{2}$ at $k$ points by ${\cal B}_{k}$.
\onefigure{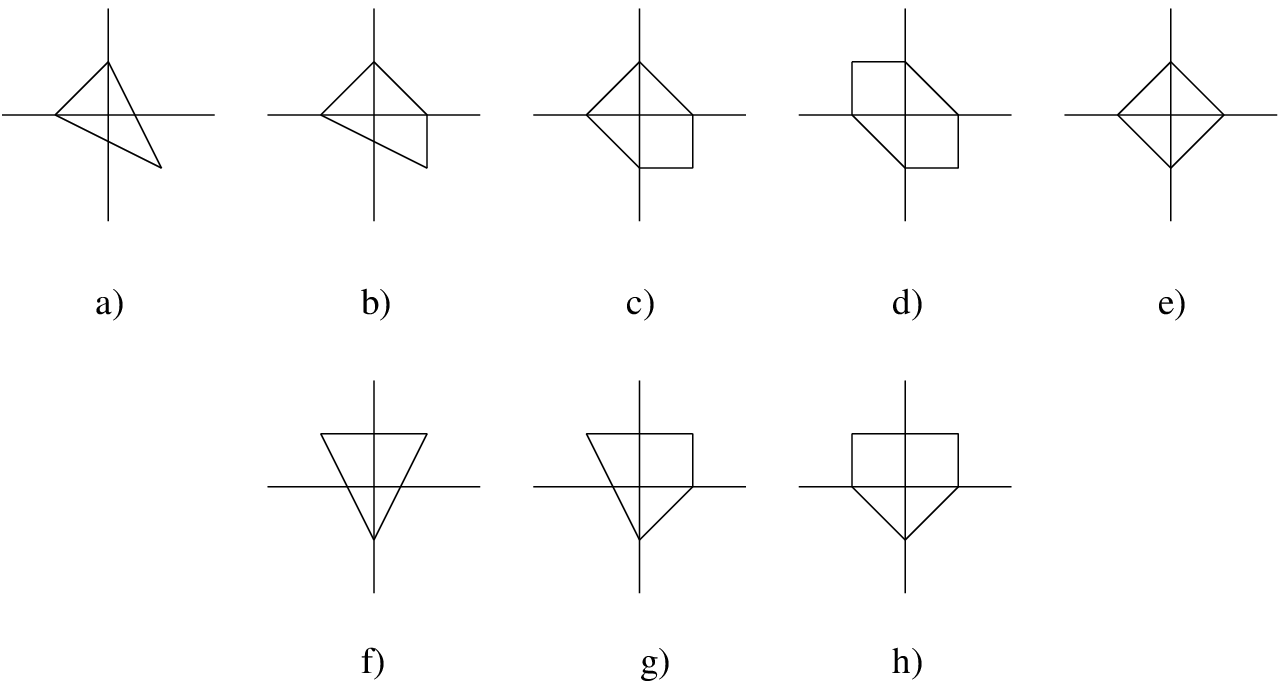}{a) $\PP^{2}$,\,\, b)${\cal B}_{1}$,\,\, c)${\cal B}_{2}$,\,\,
d)${\cal B}_{3}$,\,\, e) $\PP^{1}\times \PP^{1}$,\,\, f) $F_{2}$, g) $F_{2}$ blown up at one point,
\,\,h) $F_{2}$ blown up at two points.}
\subsection{$\PP^{2}$}

The amplitude associated with the web diagram can be written down completely
once the gluing matrices associated with each vertex are known.
All Riemann surfaces corresponding to the 5-brane webs can be
constructed from a "propagator" (a cylinder) and $\sl2z$ transform
of a "three point vertex" ( a sphere with three disks removed).
The $\sl2z$ gluing matrices associated with the vertices of the
brane web contain the information about how the propagators and
the three point vertex are to be put together to get the web.
These gluing matrices are $\sl2z$ matrices which transform one
edge of the three point vertex with charge $(p,q)$ into another
edge with charge $(r,s)$. For the case of non-compact Calabi-Yaus
with a single compact divisor the Riemann surface is of genus one
which is reflected by the fact that the 5-brane web has a single
``face''. In this case we see that each three point vertex used to
construct the web has two edges which are the internal edges and
therefore it is natural to define the gluing matrix at that vertex
to be the $\sl2z$ matrix which maps one internal edge to another
internal edge associated with that vertex consistent with the vertex defined before.

We begin with the case of $\PP^{2}$, which was discussed in detail in
\cite{AMV}, for completeness. The non-compact Calabi-Yau manifold is
the total space of ${\cal O}(-3)$ bundle over $\PP^{2}$. The web
diagram dual to this non-compact CY can be obtained from the toric
diagrams given in \figref{toric}(a) and is shown in \figref{B0}.
\onefigure{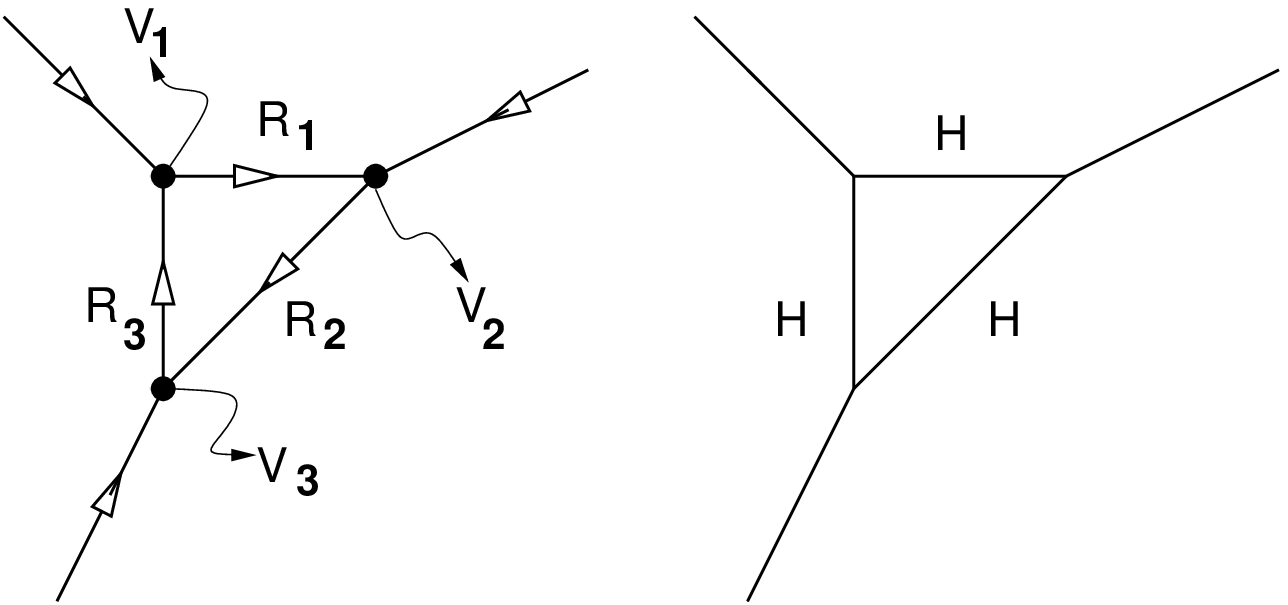}{a) The 5-brane web dual to the local $\PP^{2}$, $(p,q)$
5-brane is oriented in the direction $(p,q)$. $R_{a}$ denote the
states propagating on the internal lines, b) the rational curve
classes of the internal edges.}  The equation for the Riemann surface
which gives the dual theory when M5-brane is wrapped on it can be
easily determined from the mirror superpotential or the toric data \cite{HV,HIV} \bea
1+e^{u}+e^{v}+e^{-t-u-v}=0\,.  \eea Where $t$ is the K\"ahler
parameter of the generator of $H_{2}(\PP^{2},\bbbz)$ which we will
denote by $H$. $H$ is the class of $\PP^{1}$ given by a linear
equation in $\PP^{2}$ and is such that $H\cdot H=1$.  From the web
digram we can determine the operators associated with the
vertices. These operators are just the operators in the WZW theory
corresponding to the $\sl2z$ matrices mapping the internal edges into
each other at each vertex, \bea & &\pmatrix{0 \cr 1} \stackrel{{\cal
V}_{1}}{\mapsto} \pmatrix{1\cr 0} \stackrel{{\cal V}_{2}}{\mapsto}
\pmatrix{-1 \cr -1}\stackrel{{\cal V}_{3}}{\mapsto} \pmatrix{0\cr
1}\,,\\ \nn & &{\cal V}_{1}=S^{-1}\,,\,\,{\cal
V}_{2}=TS^{-1}T\,,\,\,{\cal V}_{3}=TS^{-1}\,.  \eea

Given these operators associated with the vertices we can write down
the instanton contribution to the closed string amplitude (this can also be
obtained as the large N limit of the Chern-Simons theory on local $B_{3}$
with three non intersecting exceptional curves geometrically
transitioned to $S^{3}$ as was done in\cite{AMV}),
\bea
e^{F^{(c)}_{inst}}&=&\sum_{R_{1},R_{2},R_{3}}e^{-\sum_{a=1}^{3}l_{a}r_{a}}
\langle \bar{R_{3}}|{\cal V}_{3}|R_{2},0\rangle \langle \bar{R_{2}}|{\cal V}_{2}|R_{1},0\rangle
\langle \bar{R_{1}}|{\cal V}_{1}|R_{3},0\rangle \,,\\ \nn
&=&\sum_{R_{1},R_{2},R_{3}}e^{-\sum_{a=1}^{3}l_{a}r_{a}}
\langle \bar{R_{3}}|{\cal V}_{3}|R_{2}\rangle \langle \bar{R_{2}}|{\cal V}_{2}|R_{1}\rangle
\langle \bar{R_{1}}|{\cal V}_{1}|R_{3}\rangle \,.
\eea

\underline{\bf Note:} Local $\PP^{2}$ blown up at more than three
points have web diagrams in which some of the external legs are either
parallel or cross each other. This is due to the fact that $\PP^{2}$
blown up at more than three points in not toric although we can choose
special points and blow them up. The problem of intersecting external
legs was solved in \cite{DHIK} by placing appropriate (p,q)7-brane and
allowing the 5-brane to end on these 7-brane so that the external
5-brane no longer can go to infinity and cross each other.  This
5-brane/7-brane picture can be simplified so that the web is given by a
single 5-brane going around the 7-brane as shown in \figref{7branes} below. 

\onefigure{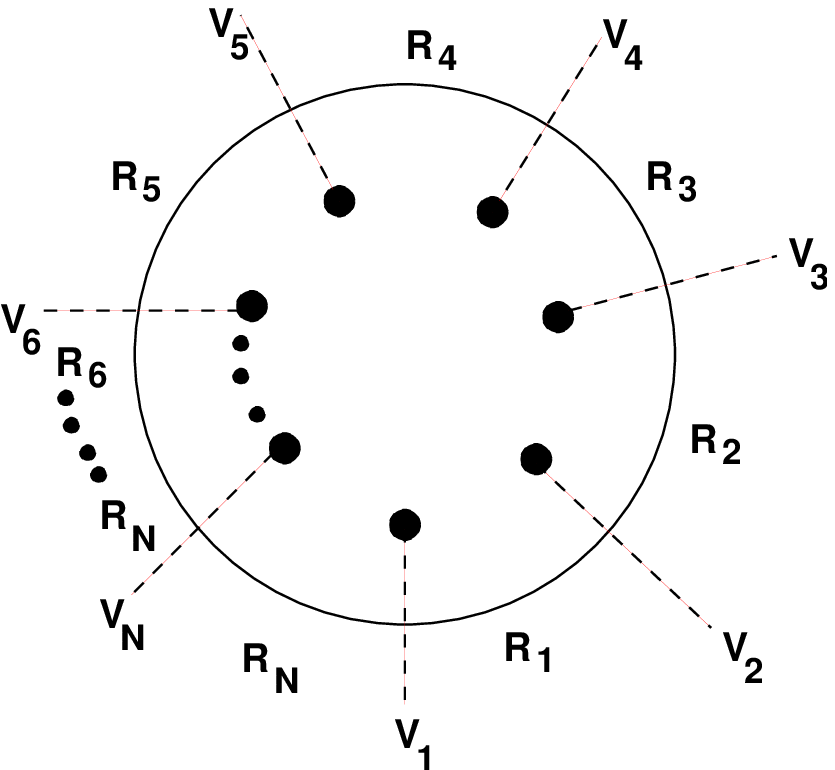}{5-brane/7-brane description of toric and non-toric local Calabi-Yau's.}

In this picture it is easy to generalize and write amplitudes for non-toric local Calabi-Yau,
\bea
e^{F^{(c)}_{inst}}:=\sum_{a=1}^{N}\sum_{R_{a}}e^{-\sum_{i=1}^{N}l_{i}r_{i}}\prod_{i=1}^{N} \langle R_{i+1}|{\cal V}_{i+1}|R_{i}\rangle \,.
\eea
Where ${\cal V}_{i}$ are the $\sl2z$ monodromy matrices associated with the 7-branes.
These amplitudes for non-toric Calabi-Yau spaces will be discussed elsewhere \cite{PI}.

Coming back to the local $\PP^{2}$ case we determine the relation
between $r_{a}$ and the renormalized K\"ahler parameter by calculating
the terms with only one non-vanishing $l_{a}$.

\underline{$l_{a}=\delta_{ab}$:} We denote the term with
$l_{a}=\delta_{ab}$ by ${\cal I}_{b}$, \bea {\cal
I}_{b}&=&e^{-r_{b}}(-1) W^{2}_{\cdot, \tableau{1}}(\lambda,q)\,,\\\nn
&=&
-e^{-r_{b}}(\frac{\lambda^{1/2}-\lambda^{-1/2}}{q^{1/2}-q^{-1/2}})^{2}\,,\\
\nn &=& -\frac{e^{-T_{b}}}{(q^{1/2}-q^{-1/2})^{2}}\,,\\ \nn &=&
\frac{e^{-T_{b}}}{(2\mbox{sin}(g_{s}/2))^{2}}\,.  \eea Where the minus
sign comes from the action of $T$ operator on the state with
fundamental representation $|\tableau{1}\rangle$ and we have defined
$r_{b}=T_{b}+\mbox{log}(\lambda)$ before taking the limit
$\lambda\mapsto \infty$.  $T_{b}$ is the "renormalized K\"ahler
parameter of the rational curve which lives on the $b-th$ internal
edge of the web diagram. Since all the rational curves living on the
internal edge of the web diagram are in the same class $H$ therefore
we can set $T_{1}=T_{2}=T_{3}=:T_{H}$.  After this redefinition we see
that we only need the leading order terms in $\lambda$ in the matrix
elements $\langle \bar{R}|S^{-1}|\bar{R'}\rangle$ as noted in
\cite{AMV}, \bea
e^{F^{(c)}_{inst}}=\sum_{R_{1},R_{2},R_{3}}e^{-(l_{1}+l_{2}+l_{3})(T_{H}-i\pi)}
{\cal W}_{R_{3},R_{2}}\,{\cal W}_{R_{2},R_{1}}\,{\cal
W}_{R_{1},R_{3}}\,q^{\frac{1}{2}(\kappa_{R_{1}}+\kappa_{R_{2}}+\kappa_{R_{3}})}\,.
\label{P2amplitude}
\eea

\subsubsection{Integer invariants of few curve classes}

We evaluate the invariants for the curve classes $H,2H$ and $3H$.
Terms contribution to $nH$ are proportional to
$e^{-nT_{H}}$ and therefore $l_{1}+l_{2}+l_{3}=n$. For $n=1$ 
it follows form Eq(\ref{P2amplitude}) that 
\bea
{\cal I}(H)=-3e^{-T_{H}}\,{\cal W}^{2}_{\tableau{1}}\,.
\eea

For $n=2$ we have  from Eq(\ref{P2amplitude}),
\bea
{\cal I}(2H)&=&e^{-2T_{H}}(3q{\cal W}^{2}_{\tableau{2}}+3q^{-1}{\cal W}^{2}_{\tableau{1 1}}
+3{\cal W}^{2}_{\tableau{1}}\,{\cal W}_{\tableau{1},\tableau{1}})\,,\\
&=& e^{-2T_{H}}\,3q^{-1}(2(q^{2}+q^{-2})+(q+q^{-1}))\,{cal W}^{2}_{\tableau{2}}\,.
\eea

For $n=3$,
\bea \nn
{\cal I}(3H)&=&-e^{-3T_{H}}(3q^{3}{\cal W}^{2}_{\tableau{3}}+3q^{-3}{\cal W}^{2}_{\tableau{1 1 1}}+3{\cal W}^{2}_{\tableau{2 1}}
+\\ \nn
&&6q{\cal W}_{\tableau{1}}{\cal W}_{\tableau{2}}{\cal W}_{\tableau{1},\tableau{2}}+
6q^{-1}{\cal W}_{\tableau{1}}{\cal W}_{\tableau{1 1}}{\cal W}_{\tableau{1},\tableau{1 1}}+{\cal W}^{2}_{\tableau{1},\tableau{1}})\\ \nn
&=&-e^{-3T_{H}}\,q^{-3}P(q)\,{\cal W}^{2}_{\tableau{3}}\,,\\ \nn
P(q)&:=&10(q^{6}+q^{-6})+7(q^{5}+q^{-5})+8(q^{4}+q^{-4})
+3(q^3+q^{-3})\\ \nn
&&+10(q^2+q^{-2})+26(q+q^{-1})+34
\eea

Thus up to $e^{-3T}$ the closed string expansion is given by \bea
F^{(c)}_{inst}&=&\mbox{log}(1+{\cal I}(H)+{\cal I}(2H)+{\cal
I}(3H)+\cdots)\,\\ \nn &=& {\cal I}(H)+\{{\cal I}(2H)-\frac{1}{2}{\cal
I}(H)^{2}\}+\{{\cal I}(3H)-{\cal I}(H){\cal I}(2H)+\frac{1}{3}{\cal
I}(H)^{3}\}+\cdots\,,\\ \nn
&=&e^{-T_{H}}\frac{3}{(2\mbox{sin}(g_{s}/2))^{2}}+
e^{-2T_{H}}\{\frac{3}{2(2\mbox{sin}(g_{s}))^{2}}-\frac{6}{(2
\mbox{sin}(g_{s}/2))^{2}}\}+e^{-3T_{H}}\{\frac{3}{3(\mbox{sin}(3g_{s}/2))^{2}}
\\ \nn &&+\frac{27}{(2\mbox{sin}(g_{s}/2))^{2}}-10\}+\cdots \eea To
get the integer invariant we compare the above expansion with the
close string expansion \bea \nn
F^{(c)}_{inst}&=&e^{-T_{H}}\frac{N^{0}_{H}}{(2\mbox{sin}(g_{s}/2))^{2}}+
e^{-2T_{H}}\{\frac{N^{0}_{H}}{2(2\mbox{sin}(g_{s}))^{2}}+\frac{N^{0}_{2H}}{(2
\mbox{sin}(g_{s}/2))^{2}}\}+e^{-3T_{H}}\{\frac{N^{0}_{H}}{3(\mbox{sin}(3g_{s}/2))^{2}}
\\ \nn
&&+\frac{N^{0}_{3H}}{(2\mbox{sin}(g_{s}/2))^{2}}+N^{1}_{3H}\}+\cdots\,,
\eea to obtain$ N^{g}_{H}=3\,\delta_{g,0}, N^{g}_{2H}=-6\delta_{g,0}$
and $N^{g}_{3H}=27\delta_{g,0}-10\delta_{g,1}$.

\subsection{local ${\cal B}_{1}$}
Consider now the case of a non-compact Calabi-Yau with a compact
divisor which is $\PP^{2}$ blown up at one point, ${\cal B}_{1}$.  The
case when the divisor is a $\PP^{2}$ can also be obtained from this as
we will show later. We will see that the geometry of the web (or the
Riemann surface given by mirror symmetry) when interpreted as a
Feynman diagram completely determines instanton corrections to the
closed topological string amplitude.

The geometry of the web corresponding to local ${\cal B}_{1}$ is shown
in \figref{B1}.  \onefigure{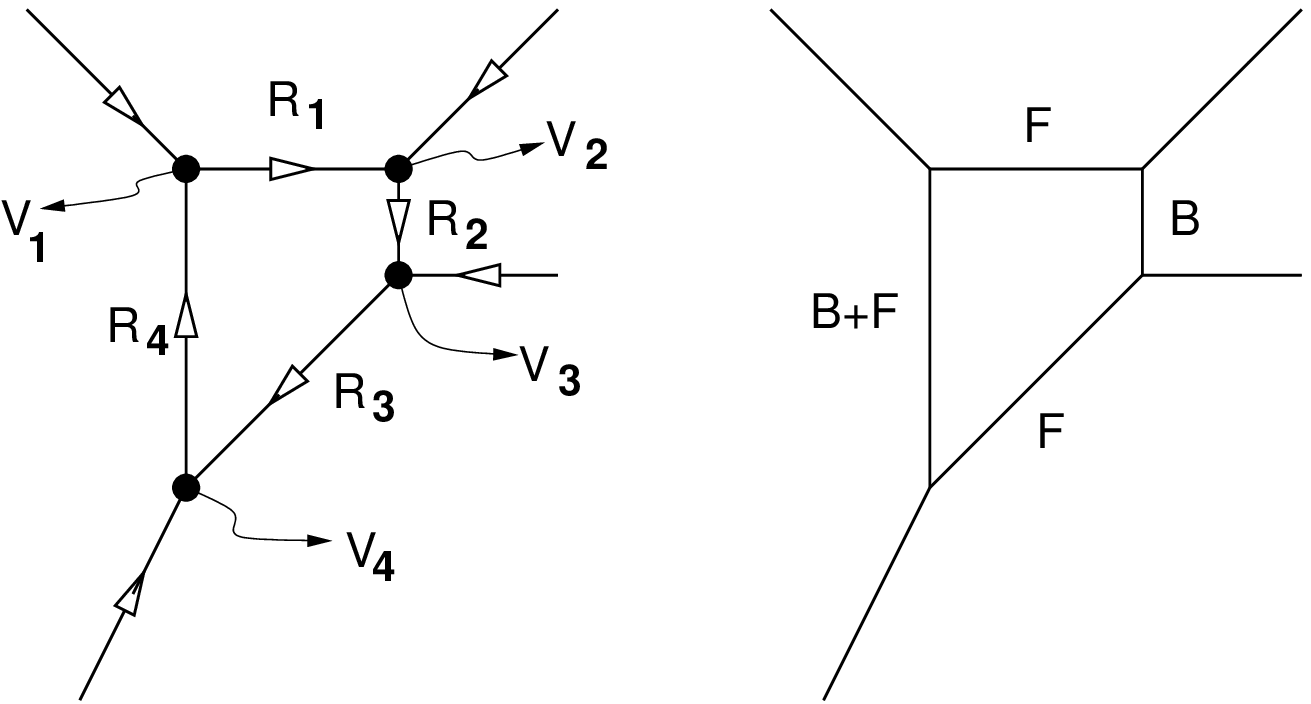}{The 5-brane web dual to the local
${\cal B}_{1}$, b) rational curves associated with the internal
edges.}  The Riemann surface can be obtained by thickening the web and
is given by the following equation, \bea
1+e^{u}+e^{v}+e^{-t_{1}-u+v}+e^{-t_{2}-u}=0\,. \eea This equation as
well as the 5-brane web can be easily obtained from the toric data shown in \figref{toric}.

\subsubsection{K\"ahler parameters and homology}
${\cal B}_{1}$ is $\PP^{2}$ blown up at one point and therefore has
$\mbox{dim}H_{2}({\cal B}_{1})=2$. As a basis we take the  curves  given
by $B:=E$ and $F:=H-E$ where $H$ is the class coming from
$\PP^{2}$ and $E$ is the class of the exceptional curve obtained
by blowing up. The intersection matrix is then , \bea
B^{2}=-1\,,\,\,B\cdot F=1\,,\,\,F^{2}=0\,. \eea We have used the
fact that ${\cal B}_{1}$ is a $\PP^{1}$ bundle over $\PP^{1}$ to use 
as basis the base $\PP^{1}$ and the fiber $\PP^{1}$ denoted by $B$
and $F$ respectively. We denote by $t_{B,F}$ the K\"ahler
parameters of $B$ and $F$. Since intersection numbers between
effective curves must be non-negative it follows that for $nB+mF$
to be effective and therefore contribute to the topological string
amplitudes we must have \bea (nB+mF)\cdot B\geq 0 \,\,\mapsto
\,\,m\geq n\,\,\,\mbox{if}\,\,m\neq 0 \eea Each internal line of the web corresponds to a
curve in ${\cal B}_{1}$. Denoting these curves by $C_{a}$  it follows
that these curves are not independent ($-K_{{\cal B}_{1}}$ is the class dual to the first Chern class of ${\cal B}_{1}$)
\bea
\sum_{a=1}^{4}C_{a}=-K_{{\cal B}_{1}}=2B+3F\,.\eea It the follows that
\bea C_{1}=F\,,\,\,C_{2}=B\,,\,\,C_{3}=F\,,\,\,C_{4}=B+F\,. 
\label{curvesB1}
\eea

From the web diagram, \figref{B1}, it is  easy to write down the
gluing matrices ${\cal V}_{a}$ associated with the vertices, \bea
\pmatrix{0\cr 1}\stackrel{{\cal V}_{1}}{\mapsto} \pmatrix{1\cr 0}
\stackrel{{\cal V}_{2}}{\mapsto}\pmatrix{0\cr -1} \stackrel{{\cal V}_{3}}{\mapsto}
\pmatrix{-1\cr -1}\stackrel{{\cal V}_{4}}{\mapsto}\pmatrix{0\cr 1}\,,\\ \nn
{\cal V}_{1}=S^{-1}\,,\,{\cal
V}_{2}=S^{-1}\,,\,{\cal V}_{3}=S^{-1}T^{-1}\,,\,{\cal
V}_{4}=TS^{-1}\,. \eea

The amplitude associated with the above 5-brane web is obtained by
interpreting the corresponding Riemann surface/web as a one loop
Feynman diagram with four external states. We take the states
associated with the external legs to be vacuum states
$|0\rangle$. And the states $|R_{a}\rangle$ propagating in the
loop are summed over. With this prescription the amplitude whose log
gives the topological closed string instanton corrections is given by
 \bea e^{F^{(c)}_{inst}}&:=&
\sum_{R_{1},R_{2},R_{3},R_{4}}e^{-\sum_{a=1}^{4}l_{R_{a}}r_{a}}\langle
\bar{R_{4}}|{\cal V}_{4}|R_{3},0\rangle \langle \bar{R_{3}}|{\cal
V}_{3}|R_{2},0\rangle \langle \bar{R_{2}}|{\cal
V}_{2}|R_{1},0\rangle \langle \bar{R_{1}}|{\cal
V}_{1}|0,R_{4}\rangle\,.\nn \,\\ \nn
&=&\sum_{R_{1},R_{2},R_{3},R_{4}}e^{-\sum_{a=1}^{4}l_{R_{a}}r_{a}}\langle
\bar{R_{4}}|{\cal V}_{4}|R_{3}\rangle \langle \bar{R_{3}}|{\cal
V}_{3}|R_{2}\rangle \langle \bar{R_{2}}|{\cal
V}_{2}|R_{1}\rangle \langle \bar{R_{1}}|{\cal
V}_{1}|R_{4}\rangle\,.\label{B1amplitude} \nn
 \eea

In the above equation $r_{a}$ is the length of the internal edge,
$l_{a}$ is the number of boxes in the Young Tableau corresponding
to the representation $R_{a}$ and  the term $e^{-l_{a}r_{a}}$
comes from the propagator joining the states at two adjacent
vertices. Note that not all $r_{a}$ are actually independent as we
will see below. Note that if $r_{2}=0$ we can sum over $R_{2}$ and
obtain the expression for the local $\PP^{2}$ case \cite{AMV}.

To determine the relation between $r_{a}$ and the renormalized
K\"ahler parameters $T_{B,F}$ we evaluate terms in the
Eq(\ref{B1amplitude}) with only one non-vanishing $l_{a}$.

\underline{$l_{a}=\delta_{ab}$:} We denote this term by $I_{b}$, \bea
I_{b}&:=&(-1)^{\delta_{b,2}+\delta_{b,4}}\,e^{-r_{b}}\,
W^{2}_{\tableau{1}}(q,\lambda)\,,\\ \nn
&=&(-1)^{\delta_{b,2}+\delta_{b,4}}\,e^{-r_{b}}\,(\frac{\lambda^{1/2}-\lambda^{-1/2}}{q^{1/2}-q^{-1/2}})^{2}\,.
\eea In order to obtain a finite non-vanishing $I_{b}$ in the limit $\lambda\mapsto \infty$ we define
the renormalized K\"ahler parameter $T_{b}$ associated with curves $C_{b}$ as follows 
\bea 
T_{b}:=r_{b}-\mbox{log}(\lambda)\,.
\eea
Thus from the above equation and Eq(\ref{curvesB1}) we get
\bea
r_{1}&=&T_{1}+\mbox{log}(\lambda)=T_{F}+\mbox{log}(\lambda)\,,\\ \nn
r_{2}&=&T_{2}+\mbox{log}(\lambda)=T_{B}+\mbox{log}(\lambda)\,,\\ \nn
r_{3}&=&T_{3}+\mbox{log}(\lambda)=T_{F}+\mbox{log}(\lambda)\,,\\ \nn
r_{4}&=&T_{4}+\mbox{log}(\lambda)=T_{B}+T_{F}+\mbox{log}(\lambda)\,.
\eea

\subsubsection{Integer invariants of few curve classes}
Using the above definition of $r_{a}$ in terms of $T_{B,F}$ we can write 
Eq(\ref{B1amplitude}) as 

\bea\nn
e^{F^{(c)}_{inst}}=\sum_{a=1}^{4}\sum_{R_{a}}\,e^{-l_{B}T_{B}-l_{F}T_{F}} {\cal W}_{R_{4},R_{3}}\,{\cal W}_{R_{3},R_{2}}\,{\cal W}_{R_{2},R_{1}}\,{\cal W}_{R_{1},R_{4}}\,(-1)^{l_{2}+l_{4}}\,q^{\frac{1}{2}(\kappa_{R_{4}}-\kappa_{R_{2}})}\,.
\label{B1amplitude2}
\eea
where
\bea
l_{B}&:=&l_{2}+l_{4}\,,\,\,l_{F}:=l_{1}+l_{3}+l_{4}\,,
\eea 
and ${\cal W}_{R_{a},R_{b}}$ are the coefficient of the leading
power of $\lambda$, $\lambda^{\frac{l_{a}+l_{b}}{2}}$, in $\langle
R_{a}|S^{-1}|R_{b}\rangle$ and some of them are listed in section
3. We denote the term with $e^{-mT_{F}-nT_{B}}$ in the above expansion by
${\cal I}(nB+mF)$. Using Eq(\ref{B1amplitude2}) we evaluate the
integer invariants associated with $B$, $F$ and $B+F$. 

\underline{$B$:}\,\,To obtain the coefficient of $e^{-T_{B}}$ we see
that we must have $l_{1}=l_{3}=l_{4}=0$ and $l_{2}=1$, \bea {\cal
I}(B)&=& -e^{-T_{B}}\,{\cal W}^{2}_{\tableau{1}}\,,\\ \nn
&=&-\frac{e^{-T_{B}}}{(q^{1/2}-q^{-1/2})^{2}}\,,\\ \nn
&=&\frac{e^{-T_{B}}}{(2\mbox{sin}(g_{s}/2))^{2}}\,,\eea where
$\lambda=q^N$ and $q=e^{ig_{s}}$.  Since this curve class is not a
multiple of another curve class therefore we do not have to take into
account any multiple cover contributions and comparing it with the
closed string expansion gives \bea
N^{0}_{B}=1\,,\,\,N^{g}_{B}=0\,,\,g>0\,.\eea Which is indeed the
correct result given that the exceptional curve $B$ is rigid and has
zero dimensional moduli space.

\underline{$F$:}\,\, In this case we must have
$(l_{1},l_{2},l_{3},l_{4})=(1,0,0,0)\,\, \mbox{or}\,\, (0,0,1,0)$.
Again denoting the term by ${\cal I}(F)$ we get \bea {\cal
I}(F)&=&\,e^{-T_{F}}\,2\,{\cal W}^{2}_{\tableau{1}}\, ,\\ \nn &=&
-\frac{2\,e^{-T_{F}}}{(2\mbox{sin}(g_{s}/2))^{2}}\,.\eea Again since
there are no multiple cover contribution to be subtracted comparing 
with the closed string expansion gives \bea
N^{0}_{F}=-2\,,\,\,\,N^{g}_{F}=0\,\,\,g>0\,. \eea Which is indeed the
correct result, since its moduli space is just the base curve which is $\PP^{1}$ 
(since $F$ is genus zero the moduli space of flat connections is trivial)
and therefore $N^{0}_{F}=(-1)^{\mbox{dim}{\cal
M}_{F}}\chi({\cal M}_{F})=-2$.

\underline{$B+F$:}\,\,Terms contributing in this case are the ones with
$(l_{1},l_{2},l_{3},l_{4})=(1,1,0,0),(0,1,1,0)$ and $(0,0,0,1)$. \bea
{\cal I}(B+F)&=&
e^{-T_{B}-T_{F}}\{-2\,{\cal W}^{2}_{\tableau{1}}\,{\cal W}_{\tableau{1},\tableau{1}}-{\cal W}^{2}_{\tableau{1}}\}\,,\\ \nn
&=&\frac{3\,e^{-T_{B}-T_{F}}}{(2\mbox{sin}(g_{s}/2))^{2}}-\frac{2\,e^{-T_{B}-T_{F}}}
{(2\mbox{sin}(g_{s}/2))^{4}}\,.
\eea 

Thus we get 
\bea F^{(c)}_{inst}&=&\mbox{log}(1+{\cal I}(B)+{\cal I}(F)+{\cal
I}(B+F)+\cdots)\,,\\ \nn &=& {\cal I}(B)+{\cal I}(F)+{\cal
I}(B+F)-{\cal I}(B){\cal I}(F)+\cdots \,,\\ \nn &=&
\frac{e^{-T_{B}}}{(2\mbox{sin}(g_{s}/2))^{2}}-\frac{2e^{-T_{F}}}{(2\mbox{sin}(g_{s}/2))^{2}}+\frac{3e^{-T_{B}-T_{F}}}{(2\mbox{sin}(g_{s}/2))^{2}}+\cdots
\eea
Thus we get 
\bea
N^{0}_{B+F}=3\,,\,\,N^{g}_{B+F}=0\,,\,\,g>0\,.
\eea
Which is indeed correct since $B+F=H$.

\subsection{${\cal B}_{2}$}

${\cal B}_{2}$ is $\PP^{2}$ blown-up at two points and can also be
obtained from $\PP^{1}\times \PP^{1}$ by blowing up one point.  The
5-brane web diagram can be obtained from the toric diagram \ref{toric}
directly or can be obtained from the mirror Riemann surface \bea
1+e^{u}+e^{v}+e^{-t_{1}-u-v}+e^{-t_{2}-u}+e^{-t_{3}-v}\,, \eea which
is embedded in $\mbox{R}^{2}\times T^{2}$ by taking the K\"ahler parameter of
$T^{2}$ to zero. Both these description give the same 5-brane web shown 
in \figref{B2} below.
\onefigure{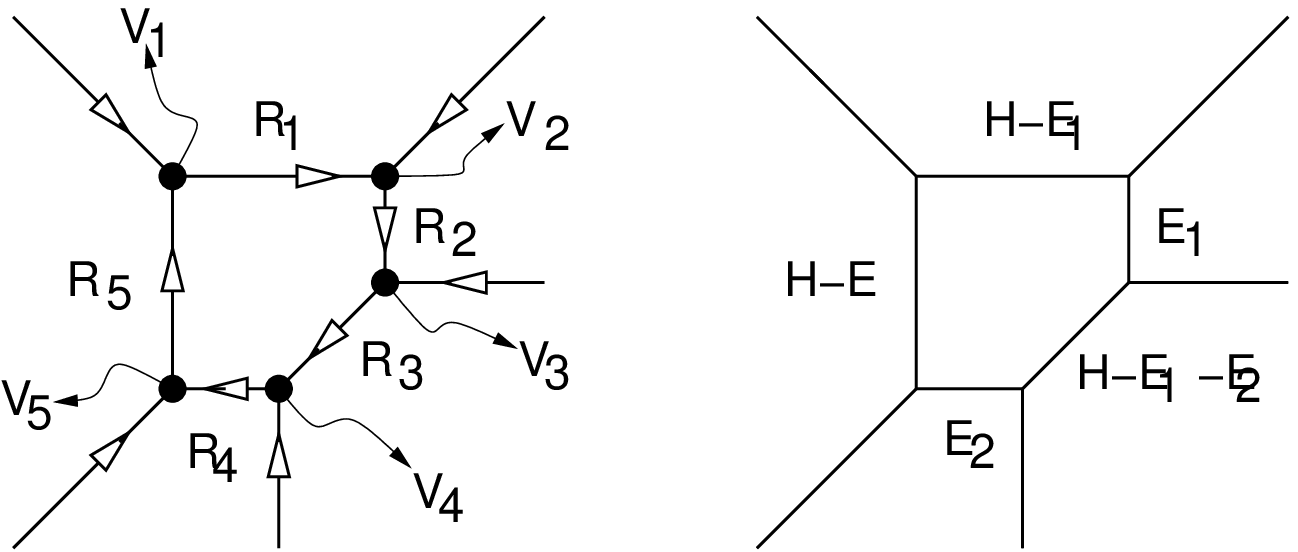}{5-brane web dual to local $B_{2}$.}
From the toric diagram we  immediately obtain the $\sl2z$ matrices
associated with the trivalent vertices,
\bea
\pmatrix{0\cr 1}\stackrel{{\cal V}_{1}}{\mapsto} \pmatrix{1\cr 0}
\stackrel{{\cal V}_{2}}{\mapsto}\pmatrix{0\cr -1} \stackrel{{\cal V}_{3}}{\mapsto}
\pmatrix{-1\cr -1}\stackrel{{\cal V}_{4}}{\mapsto}\pmatrix{-1\cr 0}\stackrel{{\cal V}_{5}}\mapsto \pmatrix{0\cr 1}\,,\\ \nn
{\cal V}_{1}=S^{-1}\,,\,{\cal
V}_{2}=S^{-1}\,,\,{\cal V}_{3}=S^{-1}T^{-1}\,,\,{\cal
V}_{4}=T^{-1}S^{-1}T^{-1}\,,\,{\cal V}_{5}=S^{-1}\,. \eea

Using these matrices we can immediately write down the amplitude
associated with the web diagram, \bea
e^{F^{(c)}_{inst}}:=\sum_{a=1}^{5}\sum_{R_{a}}e^{-\sum_{b=1}^{5}l_{b}r_{b}}
\prod_{i=1}^{5}\langle R_{i+1}|{\cal V}_{i+1}|R_{i}\rangle \,, 
\label{B2amplitude}
\eea
where $R_{6}=R_{1}$ and ${\cal V}_{6}={\cal V}_{1}$. $r_{b}$ are the
bare K\"ahler parameters associated with the curve classes $C_{b}$
forming the internal line of the web,
\bea
C_{1}=H-E_{1}\,,\,C_{2}=E_{1}\,,\,C_{3}=H-E_{1}-E_{2}\,,\,\,C_{4}=E_{2}\,,\,\,
C_{5}=H-E_{2}\,.
\eea

To determine the relation between $r_{b}$ and the renormalized
parameters associated with $C_{b}$ denoted by $T_{b}$ we calculate the terms in
Eq(\ref{B2amplitude}) with $l_{a}=\delta_{ab}$.

\underline{$l_{a}=\delta_{ab}$:} we denote this term by $I_{b}$, \bea
I_{b}&=&(-1)^{\delta_{b,2}+\delta_{b,3}+\delta_{b,4}}\,e^{-r_{b}}\,W^{2}_{\tableau{1}}\,,\\
\nn
&=&(-1)^{\delta_{b,2}+\delta_{b,3}+\delta_{b,4}}\,e^{-r_{b}}\,(\frac{\lambda^{1/2}-\lambda^{-1/2}}{q^{1/2}-q^{-1/2}})^{2}\,.
\eea Thus to be able to take the limit $\lambda \mapsto \infty$ we
define the $r_{b}$ in terms of the renormalized K\"ahler parameters
$T_{H}, T_{E_{1}}$ and $T_{E_{2}}$, 
\bea
r_{1}&=&T_{1}+\mbox{log}(\lambda)=T_{H}-T_{E_{1}}+\mbox{log}(\lambda)\,,\\ \nn 
r_{2}&=&T_{2}+\mbox{log}(\lambda)=T_{E_{1}}+\mbox{log}(\lambda)\,,\\ \nn
r_{3}&=&T_{3}+\mbox{log}(\lambda)=T_{H}-T_{E_{1}}-T_{E_{2}}+\mbox{log}(\lambda)\,,\\ \nn
r_{4}&=&T_{4}+\mbox{log}(\lambda)=T_{E_{2}}+\mbox{log}(\lambda)\,, \\ \nn
r_{5}&=&T_{5}+\mbox{log}(\lambda)=T_{H}-T_{E_{2}}+\mbox{log}(\lambda)\,
\eea

\subsubsection{Integer invariants of few curve classes}
Using the above definition of $T_{H,E_{1},E_{2}}$ we can write the  amplitude
Eq(\ref{B2amplitude}) as
\bea
e^{F^{(c)}_{inst}}=\sum_{a=1}^{5}\sum_{R_{a}}e^{-(l_{H}T_{H}-l_{E_{1}}T_{E_{1}}-l_{E_{2}}T_{E_{2}})}\,\,(-1)^{l_{2}+l_{3}+l_{4}}\,q^{-\frac{1}{2}(\kappa_{R_{2}}+\kappa_{R_{3}}+\kappa_{R_{4}})}\,\,\prod_{i=1}^{5} {\cal W}_{R_{i+1},R_{i}}\,.
\eea
where 
\bea
l_{H}=l_{1}+l_{3}+l_{5}\,,\,\,l_{E_{1}}=l_{1}+l_{3}-l_{2}\,,\,\,l_{E_{2}}=l_{3}+l_{5}-l_{4}\,.
\eea

we start with determining the invariants associated with the exceptional curves
$E_{1}, E_{2}$ and $H-E_{1}-E_{2}$. 

\underline{$E_{1}$:}\,\,This term, which we will denote by ${\cal
I}(E_{1})$, is given by $(l_{1},l_{2},l_{3},l_{4},l_{5})=(0,1,0,0,0)$,
\bea
{\cal I}(E_{1})&=&-e^{-T_{E_{1}}}\,{\cal W}^{2}_{\tableau{1}}\,,\\ \nn
&=& \frac{e^{-T_{E_{1}}}}{(2\mbox{sin}(g_{s}/2))^{2}}\,.
\eea
This implies that
\bea
N^{0}_{E_{1}}=1\,,\,\,\,N^{g}_{E_{1}}=0\,,\,g>0\,.
\eea

Similarly the term corresponding to $E_{2}$ is the same as above giving
$N^{0}_{E_{2}}=1$ and $N^{g}_{E_{2}}$ for $g>0$ as expected since the exceptional
curves are rigid with zero dimensional moduli space.

\underline{$H-E_{1}-E_{2}$}\,\,There is only term which contributes
here given by $(l_{1},l_{2},l_{3},l_{4},l_{5})=(0,0,1,0,0)$, \bea
{\cal I}(H-E_{1}-E_{2})&=&-e^{-(T_{H}-T_{E_{1}}-T_{E_{2}})}\,{\cal
W}^{2}_{\tableau{1}}\,,\\ \nn
&=&\frac{e^{-T_{H}+T_{E_{1}}+T_{E_{2}}}}{(2\mbox{sin}(g_{s}/2))^{2}}\,.
\eea Thus we get $N^{g}_{H-E_{1}-E_{2}}=\delta_{g,0}$ since
$H-E_{1}-E_{2}$ is an exceptional curve.

\subsection{${\cal B}_{3}$}
The web diagram in this case is given by \figref{B3}.
\onefigure{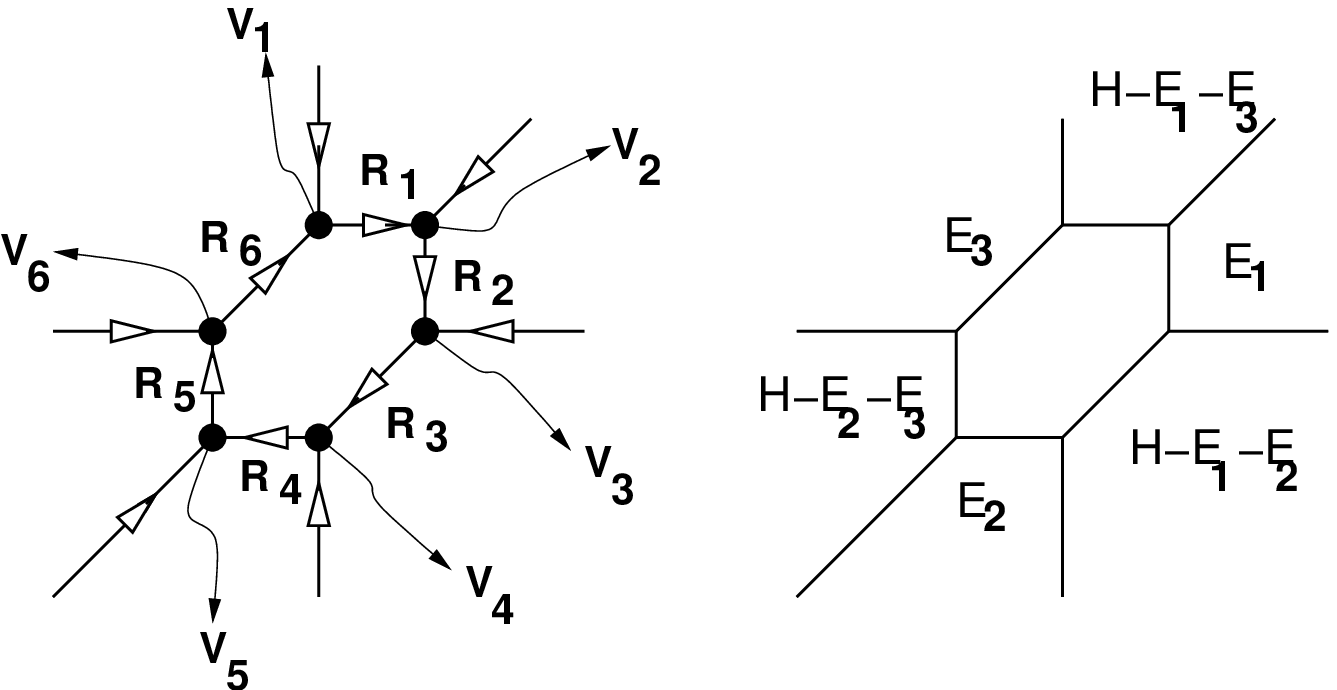}{5-brane web dual to local ${\cal B}_{3}$.}
The $\sl2z$ matrices are determined by the web diagram and are 
\bea
&&\pmatrix{1\cr 1}\stackrel{{\cal V}_{1}}{\mapsto} \pmatrix{1\cr 0}
\stackrel{{\cal V}_{2}}{\mapsto}\pmatrix{0\cr -1} \stackrel{{\cal V}_{3}}{\mapsto}
\pmatrix{-1\cr -1}\stackrel{{\cal V}_{4}}{\mapsto}\pmatrix{-1\cr 0}\stackrel{{\cal V}_{5}}\mapsto \pmatrix{0\cr 1}\stackrel{{\cal V}_{6}}{\mapsto}\pmatrix{1\cr 1}\,,\\ \nn
&&{\cal V}_{1}=T^{-1}S^{-1}T^{-1}\,,\,{\cal
V}_{2}=S^{-1}\,,\,{\cal V}_{3}=S^{-1}T^{-1}\,,\,{\cal
V}_{4}=T^{-1}S^{-1}T^{-1}\,,\,{\cal V}_{5}=S^{-1}\,,\,{\cal V}_{6}=S^{-1}T^{-1}\,. \eea

Then $F^{(c)}_{inst}$ is given by
\bea
e^{F^{(c)}_{inst}}=\sum_{a=1}^{6}\sum_{R_{a}}e^{-\sum_{b=1}^{6}l_{b}r_{b}}
\prod_{b=1}^{6}\langle R_{b+1}|{\cal V}_{b+1}|R_{b}\rangle\,.
\label{B3amplitude}
\eea
Where $R_{7}=R_{1}$ and ${\cal V}_{7}={\cal V}_{1}$.
calculation similar to the one done before gives the $r_{b}$ in terms of normalized K\"ahler
parameters $T_{H},T_{E_{1}},T_{E_{2}},T_{E_{3}}$,
\bea
r_{1}&=&T_{H}-T_{E_{1}}-T_{E_{3}}+\mbox{log}(\lambda)\,,\\ \nn
r_{2}&=&T_{E_{1}}+\mbox{log}(\lambda)\,,\\ \nn
r_{3}&=&T_{H}-T_{E_{1}}-T_{E_{2}}+\mbox{log}(\lambda)\,,\\ \nn
r_{4}&=&T_{E_{2}}+\mbox{log}(\lambda)\,,\\ \nn
r_{5}&=&T_{H}-T_{E_{2}}-T_{E_{3}}+\mbox{log}(\lambda)\,,\\ \nn
r_{6}&=&T_{E_{3}}+\mbox{log}(\lambda)\,.
\eea

\subsubsection{Integer invariants of few curve classes}
To show that this correctly reproduces the integer invariants
\cite{GV2} we calculate some terms in the above expansion
corresponding to exceptional curves. First we rewrite
Eq(\ref{B3amplitude}) in terms of $T_{H}, T_{E_{1}},T_{E_{2}},T_{E_{3}}$,
\bea
e^{F^{(c)}_{inst}}=\sum_{a=1}^{6}\sum_{R_{a}}e^{-(l_{H}T_{H}-l_{E_{1}}T_{E_{1}}-l_{E_{2}}T_{E_{2}})}\,(-1)^{\sum_{i=1}^{6}l_{i}}q^{-\frac{1}{2}(\sum_{i=1}^{6}\kappa_{R_{i}})}\,\prod_{b=1}^{6} {\cal W}_{R_{b+1},R_{b}}
\eea
Where
\bea
l_{H}=l_{1}+l_{3}+l_{5}\,,\,\,l_{E_{1}}=l_{1}+l_{3}-l_{2}\,,\,\,
l_{E_{2}}=l_{3}+l_{5}-l_{4}
\,,\,\,l_{E_{3}}=l_{1}+l_{5}-l_{6}\,.
\eea

\underline{$E_{a}$:}
In this case we have to take $l_{1}=l_{3}=l_{5}=0$ and $l_{a}=\delta_{a,2}+\delta_{a,4}=\delta_{a,6}$.
\bea
{\cal I}(E_{a})&=&-e^{-T_{E_{a}}},{\cal W}^{2}_{\tableau{1}}\,,\\ \nn
&=& \frac{e^{-T_{E_{a}}}}{(2\mbox{sin}(g_{s}/2))^{2}}\,.
\eea
Thus we get as expected $N^{g}_{E_{a}}=\delta_{g,0}$.

\underline{$H-E_{a}-E_{b}, a\neq b$:} In this case we have only one
term contributing for each curve. For $H-E_{1}-E_{3}$ we have
$l_{a}=(0,0,1,0,0,0)$, for $H-E_{1}-E_{3}$ we have
$l_{a}=(0,1,0,0,0,0)$ and for $H-E_{2}-E_{3}$ we have
$l_{a}=(1,0,0,0,0,0)$. Thus we get \bea {\cal
I}(H-E_{a}-E_{b})&=&-e^{-(T_{H}-T_{E_{a}}-T_{E_{b}})}\,{\cal
W}^{2}_{\tableau{1}}\,,\\ \nn
&=&\frac{e^{-(T_{H}-T_{E_{1}}-T_{E_{2}})}}{(2\mbox{sin}(g_{s}/2))^{2}}\,.
\eea Thus again we get the correct result
$N^{g}_{H-E_{a}-E_{b}}=\delta_{g,0}$.

\subsection{$F_{0}=\PP^{1}\times \PP^{1}$}
This case was discussed in detail in \cite{AMV}. We include this here for completeness.
The web diagram is shown in \figref{F0} from which we get the following $\sl2z$ matrices
associated with the vertices.
\onefigure{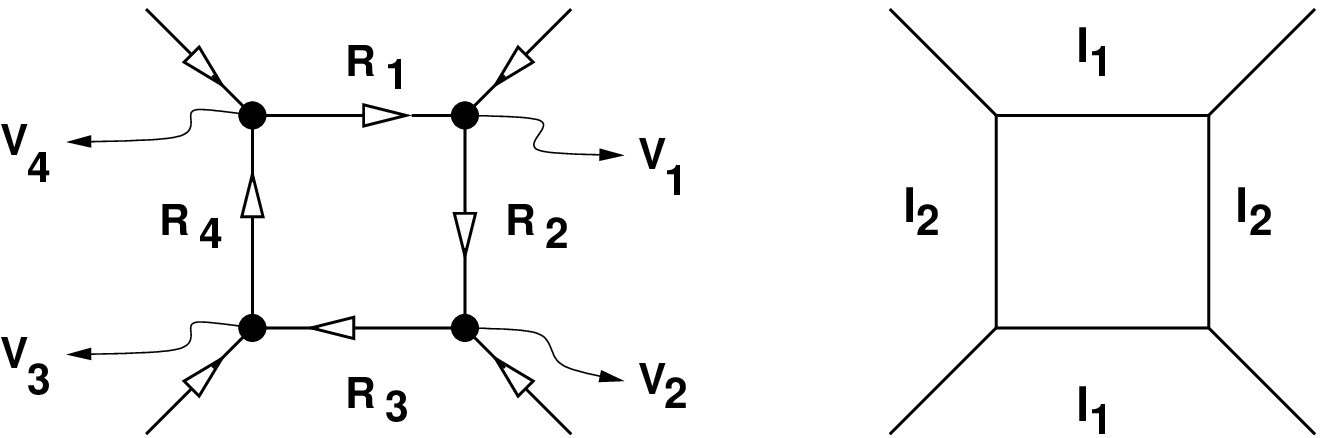}{5-brane web dual to local $\PP^{1}\times \PP^{1}$.}
\bea
&&\pmatrix{1\cr 0}\stackrel{{\cal V}_{1}}{\mapsto}\pmatrix{0\cr-1}\stackrel{{\cal V}_{2}}{\mapsto}\pmatrix{-1\cr 0}\stackrel{{\cal V}_{3}}{\mapsto}\pmatrix{0\cr 1}\stackrel{{\cal V}_{4}}{\mapsto}\pmatrix{1\cr0}\,,\\ \nn
&&{\cal V}_{1}={\cal V}_{2}={\cal V}_{3}={\cal V}_{4}=S^{-1}\,.
\eea
The amplitude associated with it is then given by 
\bea
e^{F^{(c)}_{inst}}=\sum_{a=1}^{4}\sum_{R_{a}}e^{-\sum_{a=1}^{4}l_{a}r_{a}}\,\prod_{b=1}^{4}
\langle R_{b+1}|{\cal V}_{b+1}|R_{b}\rangle.
\eea
Where we have $R_{5}=R_{1}$ and ${\cal V}_{5}={\cal V}_{1}$.
This is exactly the expression obtained in \cite{AMV} using geometric transition of $F_{0}$ blown up at four points.
The relation between the renormalized K\"ahler parameters and $r_{1,2}$ can be obtained from the terms with only
one $l_{a}$ non-vanishing.
 
\underline{$l_{a}=\delta_{ab}$}
\bea
I_{b}&=&e^{-r_{b}}W^{2}_{\tableau{1}}\,,\\ \nn
&=& e^{-r_{b}}\,(\frac{\lambda^{1/2}-\lambda^{-1/2}}{q^{1/2}-q^{-1/2}})^{2}\,.
\eea
Thus $\lambda\mapsto \infty$ limit implies that we have to define the renormalized $T_{l_{1}}$ and $T_{l_{2}}$ as follows
\bea
r_{1}=r_{3}=T_{l_{1}}+\mbox{log}(\lambda)\,,\\ \nn
r_{2}=r_{4}=T_{l_{2}}+\mbox{log}(\lambda)\,.
\eea
Here $l_{1,2}$ are the two $\PP^{1}$ such that
\bea
l_{a}\cdot l_{b}=1-\delta_{ab}\,.
\eea

\subsection{$F_{2}$}
Now consider the case of local CY manifold which has compact divisor
$F_{2}$, the second Hirzebruch surface.
We consider this cases since
it would be difficult to analyze using the geometric
transition technique \cite{AMV, DFG}. The toric diagrams are shown in
\figref{toric} form which we can obtain the the web diagram as shown in
\figref{F2}{}. In this case also to begin with we have to
determine which curve classes $C_{a}$ correspond to the four internal lines of the
web diagram.

\onefigure{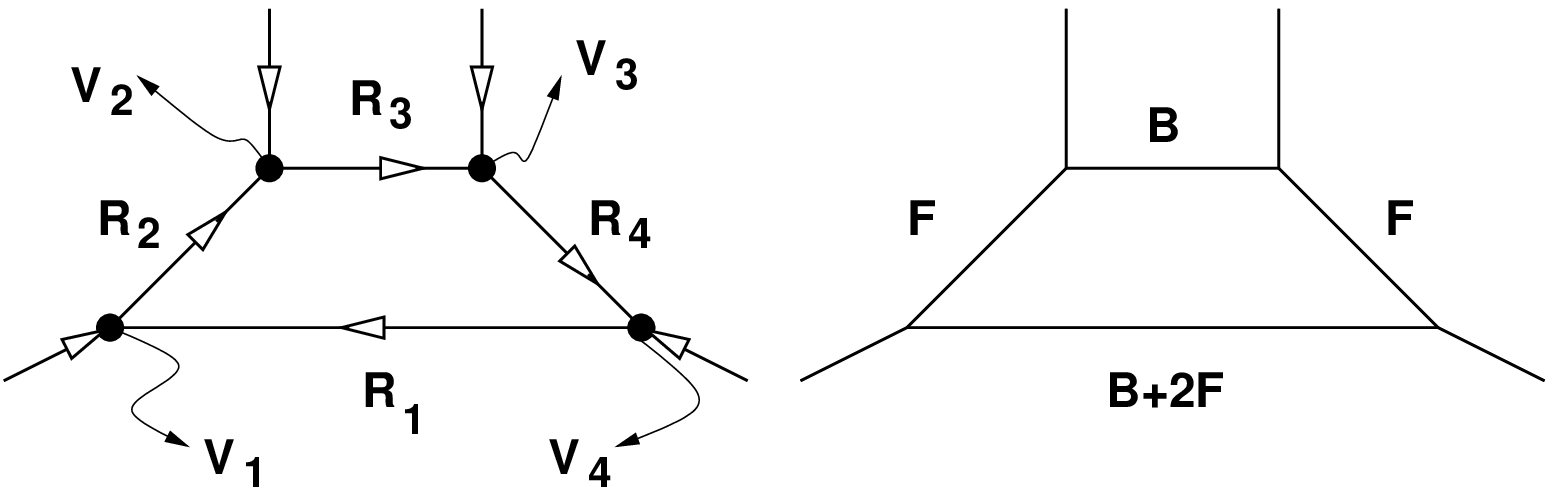}{5-brane web dual to local $F_{2}$.}          
{\bf Homology, K\"ahler parameters, gluing matrices and the Riemann surface:}
Second Hirzebruch surface $F_{2}$ like $F_{0}$ and $B_{1}=F_{1}$
is a $\PP^{1}$ bundle over $\PP^{1}$ and has a two dimensional second homology group
spanned by the class of the base $\PP^{1}$ and the class of the fiber $\PP^{1}$
which we denote with $B$ and $F$ respectively. The class dual
to the first Chern class is given by $2B+4F$ and the intersection numbers are
given by
\bea
B\cdot B=-2\,,\,\,\,B\cdot F=1\,,\,\,\,F\cdot F=0\,.
\eea
 From the web diagram it is clear that the curve between the two vertices with
 parallel external legs is a curve of self-intersection -2 and therefore
 $C_{3}=B$. Also it is clear that $C_{2}=C_{4}=F$ since $F$ is the only rational
 curve which intersects $B$ at one point.  Since $\sum_{a=1}^{4}C_{a}=2B+4F$ (class
 dual to the first Chern class) thus $C_{1}=B+2F$.

 Form \figref{F2} we can easily determine the gluing matrices as the
 $\sl2z$ matrix associated with a vertex which maps an incoming internal
 line into the outgoing internal line at that vertex. With the orientation
 of the lines as shown in \figref{F2} we get
 \bea
{\cal V}_{1}=TS^{-1}\,,\,\,{\cal V}_{2}=S^{-1}T^{-1}\,,\,\,\,
{\cal V}_{3}=T^{-1}S^{-1}
\,,\,\,\,{\cal V}_{4}=S^{-1}T\,.
 \eea
The web diagram also completely determines the equation of the Riemann surface
obtained by thickening the web. As discussed before the theory on the M5-brane wrapped
on this Riemann surface is dual to the $N=2$ theory obtained by
compactifying Type IIA string theory on the local $F_{2}$. The equation for the
Riemann surface for the local $F_{2}$ case is given by
\bea
1+e^{u}+e^{v}+e^{-t_{1}-u}+e^{-t_{2}-2u-v}=0\,.
\eea
Where $t_{1,2}$ are K\"ahler parameters and $u,v\in \bbbc$.

{\bf Topological string amplitude and renormalized K\"ahler parameter:}
Using the above gluing matrices one can immediately write down
the closed string topological amplitude by interpreting the Riemann
surface associated with the above web diagram as a Feynman digram with
operators ${\cal V}_{a}$ giving the interactions at the vertices.
\bea
e^{F_{(c)}}&=&\sum_{R_{a}}e^{-\sum_{a=1}^{4}l_{a}r_{a}}\,\langle R_{1}|{\cal V}_{4}|R_{4},0\rangle
\langle R_{4}|{\cal V}_{3}|R_{3},0\rangle\,\langle R_{3}|{\cal V}_{2}|R_{2},0\rangle
\langle R_{2}|{\cal V}_{1}|R_{1},0\rangle\,,\\ \nn
&=&\sum_{R_{a}}e^{-\sum_{a=1}^{4}l_{a}r_{a}}\,\langle R_{1}|S^{-1}T|R_{4}\rangle
\langle R_{4}|T^{-1}S^{-1}|R_{3}\rangle\,\langle R_{3}|S^{-1}T^{-1}|R_{2}\rangle
\langle R_{2}|TS^{-1}|R_{1}\rangle\,.
\label{F2amplitude}
\eea
To determine the renormalized K\"ahler parameters $T_{B}, T_{F}$ in terms of
$r_{a}$ we determine terms with
only one $l_{a}$ non-vanishing.

\underline{$l_{a}=\delta_{ab}$:}
we denote this term by ${\cal I}_{b}$ and it is the coefficient of
$e^{-r_{b}}$ in Eq(\ref{F2amplitude}),
\bea
{\cal I}_{b}&=&e^{-r_{b}}\,\langle \tableau{1}|S^{-1}|0\rangle\,\langle
0|S^{-1}|\tableau{1}\rangle\,,\\ \nn
&=&-e^{-r_{b}}(\frac{\lambda^{1/2}-\lambda^{-1/2}}{2\mbox{sin}(g_{s}/2)})^{2}\,,\\ \nn
&=&-\frac{e^{-T_{b}}}{(2\mbox{sin}(g_{s}/2))^{2}}\,.
\eea
Where in the last line we have taken the limit $\lambda\mapsto \infty$ after
defining the renormalized K\"ahler parameter $T_{C_{b}}$ associated with the curve
$C_{b}$ such that
\bea
r_{b}=T_{C_{b}}+\mbox{log}(\lambda)\,.
\eea
Thus we get the following relations between $r_{a}$ and $T_{C_{a}}$,
\bea
r_{1}&=&T_{B+2F}+\mbox{log}(\lambda)=T_{B}+2T_{F}+\mbox{log}(\lambda)\,\\ \nn
r_{2}&=&T_{F}+\mbox{log}(\lambda)\,,\\ \nn
r_{3}&=&T_{B}+\mbox{log}(\lambda)\,,\\ \nn
r_{4}&=&T_{F}+\mbox{log}(\lambda)\,.
\eea

\subsubsection{Integer invariants for few curves classes}

Using the above relations we can write Eq(\ref{F2amplitude})
in terms of $T_{B,F}$,
\bea
e^{F^{(c)}_{inst}}&=&\sum_{R_{a}}e^{-(l_{1}+l_{3})T_{B}-(2l_{1}+l_{2}+l_{4})T_{F}}
\,{\cal W}_{R_{1},R_{4}}\,{\cal W}_{R_{4},R_{3}} \,{\cal W}_{R_{3},R_{2}}
\,{\cal W}_{R_{2},R_{1}}\,.
\eea
Where as before ${\cal W}_{R_{a},R_{b}}(q)$ is the coefficient of the
leading power of $\lambda$ in $\langle \bar{\tableau{1}}|S^{-1}|0\rangle$.

We calculate the integer invariants associated with $B,F$ and $B+F$. The term
contributing to $C$ we denote by ${\cal I}(C)$, 
\bea
{\cal I}(B)&=&e^{-T_{B}}\,{\cal W}^{2}_{\tableau{1}}\,,\\ \nn
&=& -\frac{e^{-T_{B}}}{(2\mbox{sin}(g_{s}/2))^{2}}\,.
\eea
This gives $N^{g}_{B}=-\delta_{g,0}$. For $F$ we get
\bea
{\cal I}(F)&=&e^{-T_{F}}\,2{\cal W}^{2}_{\tableau{1}}\,,\\ \nn
&=&-\frac{2e^{-T_{F}}}{(2\mbox{sin}(g_{s}/2))^{2}}\,.
\eea
Which gives $N^{g}_{F}=-2\delta_{g,0}$ as expected since the moduli space of $F$ is 
just the base curve $B$ which is a $\PP^{1}$. Since we want to determine the expansion
up to $e^{-T_{B}-2T_{F}}$ therefore we need to determine also the term ${\cal I}(2F)$,
\bea
{\cal I}(2F)&=& e^{-2T_{F}}\,\{2{\cal W}^{2}_{\tableau{2}}+2{\cal W}^{2}_{\tableau{1 1}}+
{\cal W}^{4}_{\tableau{1}}\}\,,\\ \nn
&=&e^{-2T_{F}}\frac{q^{-1}+2+q}{(q^{1/2}-q^{-1/2})^{4}(q^{1/2}+q^{-1/2})^{2}}\,.
\eea
Now consider the curve $B+F$.
\bea
{\cal I}(B+F)&=&e^{-T_{B}-T_{F}}\,2{\cal W}^{2}_{\tableau{1}}\,{\cal W}_{\tableau{1},\tableau{1}}\,,\\ \nn
&=&-\frac{2\,e^{-T_{B}-T_{F}}}{(2\mbox{sin}(g_{s}/2))^{2}}+\frac{2\,e^{-T_{B}-T_{F}}}{(2\mbox{sin}(g_{s}/2))^{4}}\,.
\eea
Finally the term associated with $B+2F$ is 
\bea
{\cal I}(B+2F)&=&e^{-T_{B}-2T_{F}}\,\{{\cal W}^{2}_{\tableau{1}}+2{\cal W}_{\tableau{1}}
({\cal W}_{\tableau{1},\tableau{2}}{\cal W}_{\tableau{2}}+{\cal W}_{\tableau{1},
\tableau{1 1}}{\cal W}_{\tableau{1 1}})+{\cal W}^{2}_{\tableau{1}}{\cal W}^{2}_{\tableau{1},\tableau{1}}\}\,,\\ \nn
&=&e^{-T_{B}-2T_{F}}\,\frac{10-(q+q^{-1})-4(q^{2}+q^{-2})+4(q^{3}+q^{-3})}{(q^{1/2}-q^{-1/2})^{6}(q^{1/2}+q^{-1/2})^{2}}
\eea
\bea
F^{(c)}_{inst}&=&\mbox{log}(1+{\cal I}(B)+{\cal I}(F)+{\cal I}(2F)+{\cal I}(B+F)+{\cal I}(B+2F)+\cdots)\,,\\ \nn
&=&{\cal I}(B)+{\cal I}(F)+\{{\cal I}(2F)-\frac{1}{2}{\cal I}(F)^{2}\}+\{{\cal I}(B+F)-{\cal I}(B){\cal I}(F)\}\\ \nn
&& + \{{\cal I}(B+2F)-{\cal I}(B+F){\cal I}(F)+{\cal I}(B){\cal I}(F)^{2}-{\cal I}(B){\cal I}(2F)\}+\cdots\\ \nn
&=&-\frac{e^{-T_{B}}}{(2\mbox{sin}(g_{s}/2))^{2}}-2(\frac{e^{-T_{F}}}{(2\mbox{sin}(g_{s}/2))^{2}}+\frac{e^{-2T_{F}}}{2(2\mbox{sin}(g_{s}))^{2}})-\frac{2e^{-T_{B}-T_{F}}}{(2\mbox{sin}(g_{s}/2))^{2}}\\ \nn
&& -\frac{4e^{-T_{B}-2T_{F}}}{(2\mbox{sin}(g_{s}/2))^{2}}+\cdots
\eea
Thus we get $N^{g}_{B+F}=-2\delta_{g,0}$ and $N^{g}_{B+2F}=-4\delta_{g,0}$ as expected
\cite{KZ}.

\subsection{$F_{2}$ blown-up at one point}
The web diagram in this case is given by \figref{F2-1} from which we can immediately write
down the $\sl2z$ matrices associated with the vertices.

\onefigure{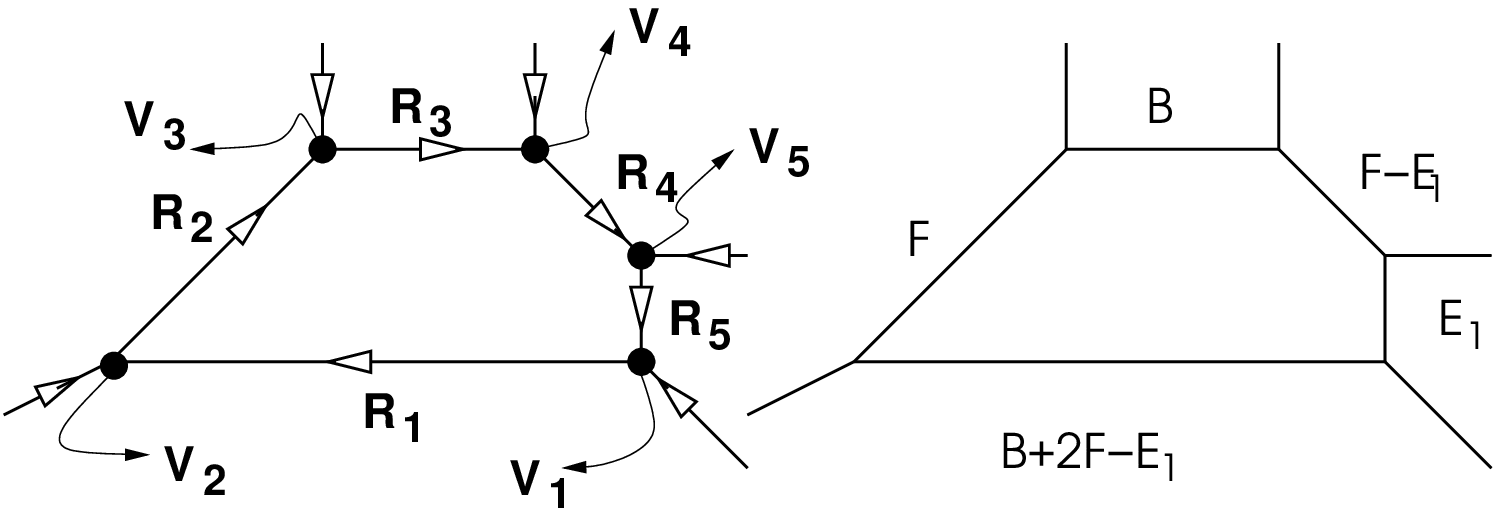}{5-brane web of the local $F_{2}$ blown-up at one point.}
\bea
{\cal V}_{1}=T^{-1}S^{-1}\,,\,\,
{\cal V}_{2}=TS^{-1}\,,\,
{\cal V}_{3}=S^{-1}T^{-1}\,,\,
{\cal V}_{4}=T^{-1}S^{-1}\,,\,
{\cal V}_{5}=TS^{-1}\,.
\eea
The amplitude is given by,
\bea
e^{F^{(c)}_{inst}}=\sum_{a=1}^{5}\sum_{R_{a}}e^{-\sum_{b=1}^{5}l_{b}r_{b}}
\prod_{b=1}^{5}\langle R_{b+1}|{\cal V}_{b+1}|R_{b}\rangle\,.
\eea
Where $R_{6}=R_{1}$ and ${\cal V}_{6}={\cal V}_{1}$.  By calculating the terms with $l_{a}=\delta_{ab}$ we can determine the renormalized; K\"ahler parameters in terms of 
$r_{b}$,
\bea
r_{1}&=&T_{B}+2T_{F}-T_{E_{1}}+\mbox{log}(\lambda)\,,\\ \nn
r_{2}&=&T_{F}+\mbox{log}(\lambda)\,,\\ \nn
r_{3}&=&T_{B}+\mbox{log}(\lambda)\,,\\ \nn
r_{4}&=&T_{F}-T_{E_{1}}+\mbox{log}(\lambda)\,,\\ \nn
r_{5}&=&T_{E_{1}}+\mbox{log}(\lambda)\,.
\eea
In terms of the renormalized K\"ahler parameters the amplitude is given by,
\bea
e^{F^{(c)}_{inst}}=\sum_{a=1}^{5}\sum_{R_{a}}\,e^{-(l_{B}T_{B}-l_{F}T_{F}-l_{E_{1}}T_{E_{1}})}\,(-1)^{l_{1}+l_{4}+l_{5}}\,q^{\frac{1}{2}(\kappa_{R_{5}}-\kappa_{R_{4}}-\kappa_{R_{1}})}\prod_{b=1}^{5}{\cal W}_{R_{b+1},R_{b}}\,.
\eea
where
\bea
l_{B}:=l_{1}+l_{3}\,,\,\,l_{F}:=2l_{1}+l_{2}+l_{4}\,,\,\,l_{E_{1}}=l_{1}+l_{4}-l_{5}\,.
\eea

We compute the invariants for exceptional curves $E_{1}$ and $F-E_{1}$
to show that the amplitude gives the correct result.

\underline{$E_{1}$:} Only one term with $(l_{1},l_{2},l_{3},l_{4},l_{5})=(0,0,0,0,1)$ contributes,
\bea
{\cal I}(E_{1})&=&e^{-T_{E_{1}}}(-1){\cal W}^{2}_{\tableau{1}}\,,\\ \nn
&=&\frac{e^{-T_{E_{1}}}{(2\mbox{sin}(g_{s}/2))^{2}}}\, \mapsto \,\,\,\,\,\, N^{g}_{E_{1}}=\delta_{g,0}\,.
\eea

\underline{$F-E_{1}$:} In this case also only one term with $l_{a}=\delta_{a,4}$ contributes,
\bea
{\cal I}(F-E_{1})&=&e^{-(T_{F}-T_{E_{1}})}\,(-1)\,{\cal W}^{2}_{\tableau{1}}\,,\\ \nn
&=&\frac{e^{-(T_{F}-T_{E_{1}})}}{(2\mbox{sin}(g_{s}/2))^{2}}\,,\,\mapsto \,\,\,\, N^{g}_{F-E_{1}}=\delta_{g,0}\,.
\eea
\subsection{$F_{2}$ blown-up at two points}
The web diagram is shown in \figref{F2-2} from which we obtain the following 
$\sl2z$ matrices associated with the vertices
\onefigure{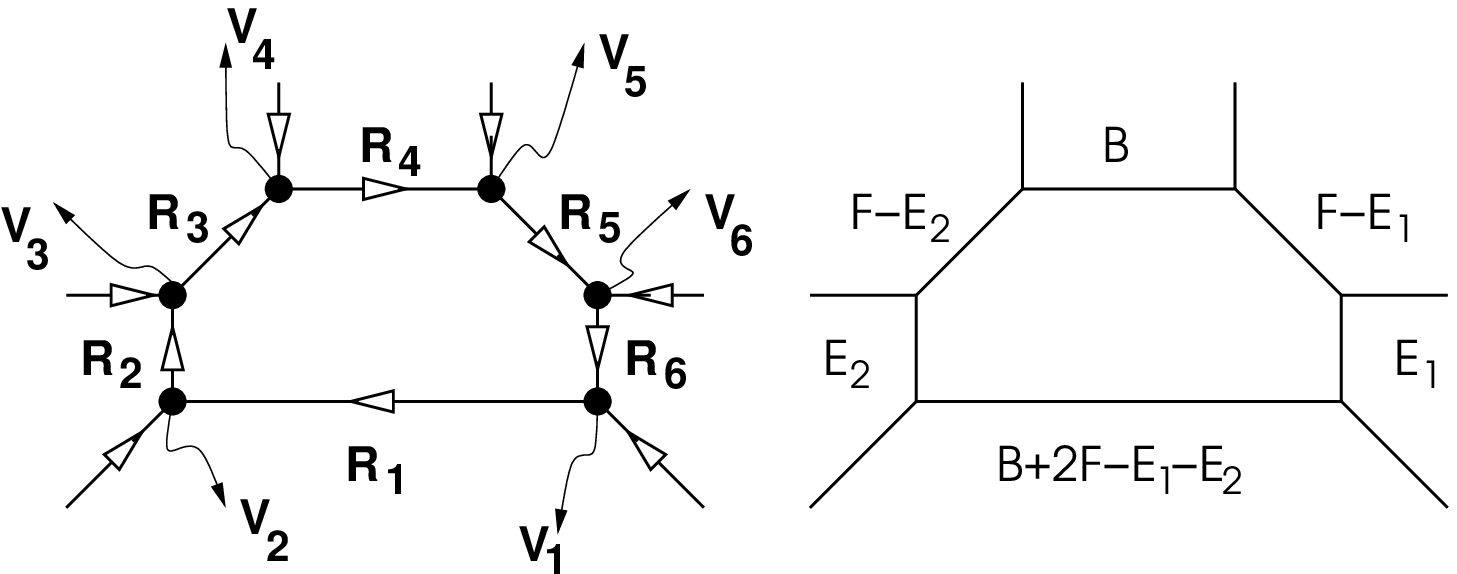}{5-brane web dual to the local $F_{2}$ blown-up at two points.}
\bea
{\cal V}_{1}=T^{-1}S^{-1}\,,\,\,{\cal V}_{2}={\cal V}_{3}={\cal V}_{4}=S^{-1}T^{-1}\,,\,\,
{\cal V}_{5}={\cal V}_{6}=T^{-1}S^{-1}\,.
\eea
The amplitude is then given by,
\bea
e^{F^{(c)}_{inst}}=\sum_{a=1}^{6}\sum_{R_{a}}e^{-\sum_{b=1}^{6}l_{b}r_{b}}
\prod_{b=1}^{6}\langle R_{b+1}|{\cal V}_{b+1}|R_{b}\rangle\,.
\eea
Where $R_{7}=R_{1}$ and ${\cal V}_{7}={\cal V}_{1}$.  By calculating the terms with $l_{a}=\delta_{ab}$ we can determine the renormalized K\"ahler parameters in terms of 
$r_{b}$,
\bea
r_{1}&=&T_{B}+2T_{F}-T_{E_{1}}-T_{E_{2}}+\mbox{log}(\lambda)\,,\\ \nn
r_{2}&=&T_{E_{2}}+\mbox{log}(\lambda)\,,\\ \nn
r_{3}&=&T_{F}-T_{E_{2}}+\mbox{log}(\lambda)\,,\\ \nn
r_{4}&=&T_{B}+\mbox{log}(\lambda)\,,\\ \nn
r_{5}&=&T_{F}-T_{E_{1}}+\mbox{log}(\lambda)\,,\\ \nn
r_{6}&=&T_{E_{1}}+\mbox{log}(\lambda)\,.
\eea
In terms of $T_{B,F,E_{1},E_{2}}$ the amplitude is given by
\bea\nn
e^{F^{(c)}_{inst}}=\sum_{a=1}^{6}\sum_{R_{a}}e^{-(l_{B}T_{B}+l_{F}T_{F}-l_{E_{1}}T_{E_{1}}-l_{E_{2}}T_{E_{2}})}\,
(-1)^{l_{2}+l_{3}+l_{5}+l_{6}}\,q^{-\frac{1}{2}(2\kappa_{R_{1}}+\kappa_{R_{2}}+\kappa_{R_{3}}+\kappa_{R_{5}}+\kappa_{R_{6}})}\,\prod_{b=1}^{6}{\cal W}_{R_{b+1},R_{b}}\,.
\eea
where
\bea
l_{B}:=l_{1}+l_{4}\,,\,\,l_{F}:=2l_{1}+l_{3}+l_{5}\,,\,\,l_{E_{1}}:=l_{1}+l_{5}-l_{6}\,\,\,l_{E_{2}}:=l_{1}+l_{3}-l_{2}\,.
\eea

\section{Conclusions}
We have seen that topological closed string amplitudes can be written
down directly from the web diagram although underlying reason for a
such a simple description of these amplitudes is still a mystery. It
would be extremely interesting to find this underlying theory giving
the propagator and the three point vertex not only because it would
give important insight into closed string amplitudes but it may also
provide a simple way of computing all genus open string amplitudes.
We believe that open string amplitudes also have a similar description
in terms of 5-brane webs with extra ingredient that we insert an
operator in the calculation of amplitude corresponding to the 3-cycle
on which the holomorphic curves can have boundaries as discussed in
\cite{AV,AKV}.

\onefigure{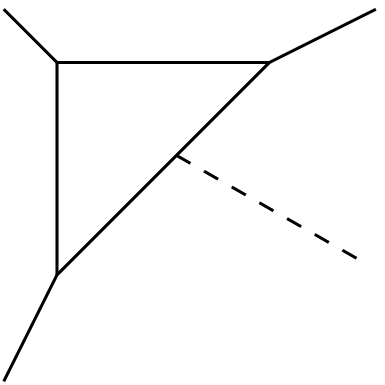}{Dashed line represents the non-compact 3-cycles 
on which D6-brane is wrapped.}

\section*{Acknowledgments}
I would like to thank Jacques Distler, Vadim Kaplunovsky and Cumrun Vafa
for valuable discussions. This research was supported by NSF
grant PHY-0071512.

\end{document}